\documentclass[aps,prb,superscriptaddress,preprint]{revtex4}
\usepackage{amsmath}
\usepackage{amssymb}
\usepackage{graphicx}
\usepackage{braket}
\usepackage{color}
\usepackage{siunitx}

\usepackage[utf8]{inputenc}
\usepackage{physics}
\usepackage{multirow}
\usepackage{url}

\begin{document}

\makeatletter
 \renewcommand\@biblabel[1]{#1.}
\makeatother

%%%%%%%%%%%%%%%%%%%%%%%%%%%%%%%%%%%%%%%%
%%%%%%% TITLE & AUTHORS
%%%%%%%%%%%%%%%%%%%%%%%%%%%%%%%%%%%%%%%%

\title{Identification and tunable optical coherent control of transition-metal spins in silicon carbide}

\author{Tom~Bosma}\thanks{These authors contributed equally to this work.}
\author{Gerrit~J.~J.~Lof}\thanks{These authors contributed equally to this work.}
\author{Carmem~M.~Gilardoni}
\author{Olger~V.~Zwier}
\author{Freddie~Hendriks}
\affiliation{Zernike Institute for Advanced Materials, University of Groningen, NL-9747AG  Groningen, The Netherlands}
\author{Bj\"orn Magnusson}
\affiliation{Department of Physics, Chemistry and Biology, Link\"oping University, SE-581 83 Link\"oping, Sweden}
\affiliation{Norstel AB, Ramsh\"allsv\"agen 15, SE-602 38 Norrk\"oping, Sweden}
\author{Alexandre Ellison}
\affiliation{Norstel AB, Ramsh\"allsv\"agen 15, SE-602 38 Norrk\"oping, Sweden}
\author{Andreas G\"allstr\"om}
\affiliation{Department of Physics, Chemistry and Biology, Link\"oping University, SE-581 83 Link\"oping, Sweden}
\affiliation{Saab Dynamics AB, SE-581 88 Linköping, Sweden}
\author{Ivan~G.~Ivanov}
\author{N.~T.~Son}
\affiliation{Department of Physics, Chemistry and Biology, Link\"oping University, SE-581 83 Link\"oping, Sweden}
\author{Remco~W.~A.~Havenith}
\affiliation{Zernike Institute for Advanced Materials, University of Groningen, NL-9747AG  Groningen, The Netherlands}
\affiliation{Stratingh Institute for Chemistry, University of Groningen, NL-9747 AG Groningen, The Netherlands}
\affiliation{Ghent Quantum Chemistry Group, Department of Inorganic and Physical Chemistry, Ghent University, B-9000 Gent, Belgium}
\author{Caspar~H.~van~der~Wal}
\affiliation{Zernike Institute for Advanced Materials, University of Groningen, NL-9747AG  Groningen, The Netherlands}

\date{Version of August 27, 2018}

% Correspondence: T. Bosma (tom.bosma@rug.nl) or C.H. van der Wal (c.h.van.der.wal@rug.nl)

\begin{abstract}
Color centers in wide-bandgap semiconductors are attractive systems for quantum technologies since they can combine long-coherent electronic spin and bright optical properties. Several suitable centers have been identified, most famously the nitrogen-vacancy defect in diamond. However, integration in communication technology is hindered by the fact that their optical transitions lie outside telecom wavelength bands. Several transition-metal impurities in silicon carbide do emit at and near telecom wavelengths, but knowledge about their spin and optical properties is incomplete. We present all-optical identification and coherent control of molybdenum-impurity spins in silicon carbide with transitions at near-infrared wavelengths. Our results identify spin $S=1/2$ for both the electronic ground and excited state, with highly anisotropic spin properties that we apply for implementing optical control of ground-state spin coherence. Our results show optical lifetimes of $\sim$60~ns and inhomogeneous spin dephasing times of $\sim$0.3~$\mu$s, establishing relevance for quantum spin-photon interfacing.
\end{abstract}

\maketitle

%%%%%%%%%%%%%%%%%%%%%%%%%%%%%%%%%%%%%%%%
%%%%%%% INTRODUCTION
%%%%%%%%%%%%%%%%%%%%%%%%%%%%%%%%%%%%%%%%

Electronic spins of lattice defects in wide-bandgap semiconductors have come forward as an important platform for quantum technologies\cite{gao2015}, in particular for applications that require both manipulation of long-coherent spin and spin-photon interfacing via bright optical transitions. In recent years this field showed strong development, with demonstrations of distribution and storage of non-local entanglement in networks for quantum communication\cite{togan2010,dolde2014,klimov2015,hensen2015,kalb2017}, and quantum-enhanced field-sensing\cite{budker2007,balasubramanian2008,dolde2011,kraus2014,bonato2016}. The nitrogen-vacancy defect in diamond is the material system that is most widely used\cite{dobrovitski2013,childress2013} and best characterized\cite{doherty2011negatively,maze2011properties,thiering2017ab}
for these applications. However, its zero-phonon-line (ZPL) transition wavelength (637~nm) is not optimal for integration in standard telecom technology, which uses near-infrared wavelength bands where losses in optical fibers are minimal. A workaround could be to convert photon energies between the emitter-resonance and telecom values\cite{radnaev2010,zaske2012,dreau2018}, but optimizing these processes is very challenging.

This situation has been driving a search for similar lattice defects that do combine favorable spin properties with bright emission directly at telecom wavelength. It was shown that both diamond and silicon carbide (SiC) can host many other spin-active color centers that could have suitable properties\cite{weber2010,weber2011,bardeleben2016,zargaleh2016} (where SiC is also an attractive material for its established position in the semiconductor device industry\cite{Friedrichs2010,song2011}). However, for many of these color centers detailed knowledge about the spin and optical properties is lacking. In SiC the divacancy\cite{koehl2011,christle2015,zwier2015} and silicon vacancy\cite{riedel2012,kraus2014,kraus2014b,widmann2015} were recently explored, and these indeed show millisecond homogeneous spin coherence times with bright ZPL transitions closer to the telecom band.

We present here a study of transition-metal impurity defects in SiC, which exist in great variety\cite{hobgood1995,baur1997,magnusson2005,ivady2011,son2012,koehl2017}. There is at least one case (the vanadium impurity) that has ZPL transitions at telecom wavelengths\cite{baur1997}, around 1300~nm, but we focus here (directed by availability of lasers in our lab) on the molybdenum impurity with ZPL transitions at 1076~nm (in 4H-SiC) and 1121~nm (in 6H-SiC), which turns out to be a highly analogous system. Theoretical investigations\cite{csore2016}, early electron paramagnetic resonance\cite{dombrowski1996,baur1997} (EPR), and photoluminescence (PL) studies\cite{kimoto1996,gallstrom2009,gallstrom2015thesis} indicate that these transition-metal impurities have promising properties. These studies show that they are deep-level defects that can be in several stable charge states, each with a distinctive value for its electronic spin $S$ and near-infrared optical transitions. Further tuning and engineering possibilities come from the fact that these impurities can be embedded in a variety of SiC polytypes (4H, 6H, etc., see Fig.~\ref{fig:FigPL}a). Recent work by Koehl~\textit{et al.}\cite{koehl2017} studied chromium impurities in 4H-SiC using optically detected magnetic resonance. They identified efficient ZPL (little phonon-sideband) emission at 1042~nm and 1070~nm, and their charge state as neutral with an electronic spin $S=1$ for the ground state.

Our work is an all-optical study of ensembles of molybdenum impurities in p-type 4H-SiC and 6H-SiC material. The charge and spin configuration of these impurities, and the defect configuration in the SiC lattice that is energetically favored, was until our work not yet identified with certainty.
Our results show that these Mo impurities are in the $\text{Mo}^{5+}$($4\text{d}^{1}$) charge state (we follow here conventional notation\cite{baur1997}: the label $5+$ indicates that of an original Mo atom 4 electrons participate in bonds with SiC and that 1 electron is transferred to the p-type lattice environment). The single remaining electron in the 4d shell gives spin $S=1/2$ for the ground state and optically excited state that we address. While we will show later that this can be concluded from our measurements, we assume it as a fact from the beginning since this simplifies the explanation of our experimental approach.

In addition to this identification of the impurity properties, we explore whether ground-state spin coherence is compatible with optical control. Using a two-laser magneto-spectroscopy method\cite{manson1994,santori2006_1,zwier2015}, we identify the spin Hamiltonian of the $S=1/2$ ground state and optically excited state, which behave as doublets with highly anisotropic Land\'e g-factors. This gives insight in how a situation with only spin-conserving transitions can be broken, and we find that we can use a weak magnetic field to enable optical transitions from both ground-state spin levels to a common excited-state level ($\Lambda$ level scheme). Upon two-laser driving of such $\Lambda$ schemes, we observe coherent population trapping (CPT, all-optical control of ground-state spin coherence and fundamental to operating quantum memories\cite{fleischhauer2005,kimble2008nature}). The observed CPT reflects inhomogeneous spin dephasing times comparable to that of the SiC divacancy\cite{zwier2015,zwierThesis2016} (near 1~$\mu$s). %yale2013,acosta2013}

In what follows, we first present our methods and results of single-laser spectroscopy performed on ensembles of Mo impurities in both SiC polytypes. Next, we discuss a two-laser method where optical spin pumping is detected. This allows for characterizing the spin sublevels in the ground and excited state, and we demonstrate how this can be extended to controlling spin coherence.

%%%%%%%%%%%%%%%%%%%%%%%%%%%%%%%%%%%%%%%%
%%%%%%% I) MATERIALS and METHODS
%%%%%%%%%%%%%%%%%%%%%%%%%%%%%%%%%%%%%%%%

\section{Materials and experimental methods}
Both the 6H-SiC and 4H-SiC (Fig.~\ref{fig:FigPL}a) samples were intentionally doped with Mo. There was no further intentional doping, but near-band-gap photoluminescence revealed that both materials had p-type characteristics.
The Mo concentrations in the 4H and 6H samples were estimated\cite{gallstrom2009,gallstrom2015thesis} to be in the range $10^{15}$-$10^{17}$~$\text{cm}^{-3}$ and $10^{14}$-$10^{16}$~$\text{cm}^{-3}$, respectively. The samples were cooled in a liquid-helium flow cryostat with optical access, which was equipped with a superconducting magnet system. The setup geometry is depicted in Fig.~\ref{fig:FigPL}b. The angle $\phi$ between the direction of the magnetic field and the c-axis of the crystal could be varied, while both of these directions were kept orthogonal to the propagation direction of excitation laser beams.
In all experiments where we resonantly addressed ZPL transitions the laser fields had linear polarization, and we always kept the direction of the linear polarization parallel to the c-axis.
Earlier studies\cite{gallstrom2009,gallstrom2015thesis,csore2016} of these materials showed that the ZPL transition dipoles are parallel to the c-axis. For our experiments we confirmed that the photoluminescence response was clearly the strongest for excitation with linear polarization parallel to the c-axis, for all directions and magnitudes of the magnetic fields that we applied.
All results presented in this work come from photoluminescence (PL) or photoluminescence-excitation (PLE) measurements. The excitation lasers were focused to a $\sim$100~$\mu$m spot in the sample.
PL emission was measured from the side. A more complete description of experimental aspects is presented in the Methods section.

%%%%%%%%%%%%%%%%%%%%%%%%%%%%%%%%%%%%%%%%%%%
%%%%%%% II) SINGLE LASER CHARACTERIZATION
%%%%%%%%%%%%%%%%%%%%%%%%%%%%%%%%%%%%%%%%%%%

\section{Single-laser characterization}
For initial characterization of Mo transitions in 6H-SiC and 4H-SiC we used PL and PLE spectroscopy (see Methods). Figure~\ref{fig:FigPL}c shows the PL emission spectrum of the 6H-SiC sample at 3.5~K, measured using an 892.7~nm laser for excitation. The ZPL transition of the Mo defect visible in this spectrum will be studied in detail throughout this work. The shaded region indicates the emission of phonon replicas related to this ZPL\cite{gallstrom2009,gallstrom2015thesis}. While we could not perform a detailed analysis, the peak area of the ZPL in comparison with that of the phonon replicas indicates that the ZPL carries clearly more than a few percent of the full PL emission. Similar PL data from Mo in the 4H-SiC sample, together with a study of the temperature dependence of the PL, can be found in the Supplementary Information (Fig.~S1).

For a more detailed study of the ZPL of the Mo defects, PLE was used. In PLE measurements, the photon energy of a narrow-linewidth excitation laser is scanned across the ZPL part of the spectrum, while resulting PL of phonon-sideband (phonon-replica) emission is detected (Fig.~\ref{fig:FigPL}b, we used filters to keep light from the excitation laser from reaching the detector, see Methods).
The inset of Fig.~\ref{fig:FigPL}c shows the resulting ZPL for Mo in 6H-SiC at 1.1057~eV (1121.3~nm). For 4H-SiC we measured the ZPL at 1.1521~eV (1076.2~nm, see Supplementary Information). Both are in close agreement with literature\cite{gallstrom2009,gallstrom2015thesis}. Temperature dependence of the PLE from the Mo defects in both 4H-SiC and 6H-SiC can be found in the Supplementary Information (Fig.~S2).

The width of the ZPL is governed by the inhomogeneous broadening of the electronic transition throughout the ensemble of Mo impurities, which is typically caused by nonuniform strain in the crystal. For Mo in 6H-SiC we observe a broadening of $24 \pm 1$~GHz FWHM, and $23 \pm 1$~GHz for 4H-SiC. This inhomogeneous broadening is larger than the anticipated electronic spin splittings\cite{baur1997}, and it thus masks signatures of spin levels in optical transitions between the ground and excited state.

%%%%%%%%%%%%%%%%%%%%%%%%%%%%%%%%%%%%%%%%%%%
%%%%%%% III) TWO-LASER CHARACTERIZATION
%%%%%%%%%%%%%%%%%%%%%%%%%%%%%%%%%%%%%%%%%%%

\section{Two-laser characterization}	
In order to characterize the spin-related fine structure of the Mo defects, a two-laser spectroscopy technique was employed\cite{santori2006_1,manson1994,zwier2015}. We introduce this for the four-level system sketched in Fig.~\ref{fig:FigSRE}a. A laser fixed at frequency $f_0$ is resonant with one possible transition from ground to excited state (for the example in Fig.~\ref{fig:FigSRE}a $\ket{g_2}$ to $\ket{e_2}$), and causes PL from a sequence of excitation and emission events. However, if the system decays from the state $\ket{e_2}$ to $\ket{g_1}$, the laser field at frequency $f_0$ is no longer resonantly driving optical excitations (the system goes dark due to optical pumping). In this situation, the PL is limited by the (typically long) lifetime of the $\ket{g_1}$ state. Addressing the system with a second laser field, in frequency detuned from the first by an amount $\delta$, counteracts optical pumping into off-resonant energy levels if the detuning $\delta$ equals the splitting $\Delta_g$ between the ground-state sublevels. Thus, for specific two-laser detuning values corresponding to the energy spacings between ground-state and excited-state sublevels the PL response of the ensemble is greatly increased. Notably, this technique gives a clear signal for sublevel splittings that are smaller than the inhomogeneous broadening of the optical transition, and the spectral features now reflect the homogeneous linewidth of optical transitions\cite{zwier2015,zwierThesis2016}.

In our measurements a 200~$\mu$W continuous-wave control and probe laser were made to overlap in the sample. For investigating Mo in 6H-SiC the control beam was tuned to the ZPL at 1121.32 nm ($f_\text{control} = f_0 = 267.3567~\text{THz})$, the probe beam was detuned from $f_0$ by a variable detuning $\delta$ (\textit{i.e.}~$f_\text{probe} = f_0 + \delta$). In addition, a $100~\mu\text{W}$ pulsed 770~nm re-pump laser was focused onto the defects to counteract bleaching of the Mo impurities due to charge-state switching\cite{beha2012,zwier2015,wolfowicz2017} (which we observed to only occur partially without re-pump laser). All three lasers were parallel to within $3^{\circ}$ inside the sample. A magnetic field was applied to ensure that the spin sublevels were at non-degenerate energies. Finally, we observed that the spectral signatures due to spin disappear in a broad background signal above a temperature of $\sim$10~K (Fig.~S4),
and we thus performed measurements at 4~K (unless stated otherwise).

Figure~\ref{fig:FigSRE}b shows the dependence of the PLE on the two-laser detuning for the 6H-SiC sample (4H-SiC data in Supplementary Information Fig.~S6), for a range of magnitudes of the magnetic field (here aligned close to parallel with the c-axis, $\phi = \ang{1}$). Two emission peaks can be distinguished, labeled line $L_1$ and $L_2$. The emission (peak height) of $L_2$ is much stronger than that of $L_1$. Figure~\ref{fig:FigSRE}c shows the results of a similar measurement with the magnetic field nearly orthogonal to the crystal c-axis ($\phi = \ang{87}$), where four spin-related emission signatures are visible, labeled as lines $L_1$ through $L_4$ (a very small peak feature left from $L_1$, at half its detuning, is an artifact that occurs due to a leakage effect in the spectral filtering that is used for beam preparation, see Methods). The two-laser detuning frequencies corresponding to all four lines emerge from the origin ($\textbf{B} = 0$, $\delta=0$) and evolve linearly with magnetic field (we checked this up to 1.2~T). The slopes of all four lines (in Hertz per Tesla) are smaller in Fig.~\ref{fig:FigSRE}c than in Fig~\ref{fig:FigSRE}b. In contrast to lines $L_1$, $L_2$ and $L_4$, which are peaks in the PLE spectrum, $L_3$ shows a dip.

In order to identify the lines at various angles $\phi$ between the magnetic field and the c-axis, we follow how each line evolves with increasing angle. Figure~\ref{fig:FigSRE}d shows that as $\phi$ increases, $L_1$, $L_3$, and $L_4$ move to the left, whereas $L_2$ moves to the right. Near $\ang{86}$, $L_2$ and $L_1$ cross. At this angle, the left-to-right order of the emission lines is swapped, justifying the assignment of $L_1$, $L_2$, $L_3$ and $L_4$ as in Fig.~\ref{fig:FigSRE}b,c. The Supplementary Information presents further results from two-laser magneto-spectroscopy at intermediate angles $\phi$ (section~2a).

%%%%%%%%%%%%%%%%%%%%%%%%%%%%%%%%%%%%%%%%
%%%%%%% IV) ANALYSIS and DISCUSSION
%%%%%%%%%%%%%%%%%%%%%%%%%%%%%%%%%%%%%%%%

\section{Analysis}
We show below that the results in Fig.~\ref{fig:FigSRE} indicate that the Mo impurities have electronic spin $S=1/2$ for the ground and excited state. This contradicts predictions and interpretations of initial results\cite{baur1997,gallstrom2009,gallstrom2015thesis,csore2016}. Theoretically, it was predicted that the defect associated with the ZPL under study here is a Mo impurity in the asymmetric split-vacancy configuration (Mo impurity asymmetrically located inside a Si-C divacancy), where it would have a spin $S = 1$ ground state with zero-field splittings of about 3 to 6~GHz\cite{baur1997,gallstrom2009,gallstrom2015thesis,csore2016}. However, this would lead to the observation of additional emission lines in our measurements. Particularly, in the presence of a zero-field splitting, we would expect to observe two-laser spectroscopy lines emerging from a nonzero detuning\cite{zwier2015}. We have measured near zero fields and up to 1.2~T, as well as from 100~MHz to 21~GHz detuning (Supplementary Information section~2c), but found no more peaks than the four present in Fig.~\ref{fig:FigSRE}c. A larger splitting would have been visible as a splitting of the ZPL in measurements as presented in the inset of Fig.~\ref{fig:FigPL}c, which was not observed in scans up to 1000~GHz. Additionally, a zero-field splitting and corresponding avoided crossings at certain magnetic fields would result in curved behavior for the positions of lines in magneto-spectroscopy. Thus, our observations rule out that there is a zero-field splitting for the ground-state and excited-state spin sublevels. In this case the effective spin-Hamiltonian\cite{abragam1970} can only take the form of a Zeeman term
\begin{equation}\label{eq:spinHam}
  H_{g(e)} = \mu_B  g_{g(e)} \mathbf{B} \cdot \tilde{\mathbf{S}}
\end{equation}
where $g_{g(e)}$ is the g-factor for the electronic ground (excited) state (both assumed positive), $\mu_B$ the Bohr magneton, $\mathbf{B}$ the magnetic field vector of an externally applied field, and  $\tilde{\mathbf{S}}$ the effective spin vector. The observation of four emission lines can be explained, in the simplest manner, by a system with spin $S = 1/2$ (doublet) in both the ground and excited state.

For such a system, Fig.~\ref{fig:FigSchemes} presents the two-laser optical pumping schemes that correspond to the observed emission lines $L_1$ through $L_4$. Addressing the system with the V-scheme excitation pathways from Fig.~\ref{fig:FigSchemes}c leads to increased pumping into a dark ground-state sublevel, since two excited states contribute to decay into the off-resonant ground-state energy level while optical excitation out of the other ground-state level is enhanced. This results in reduced emission observed as the PLE dip feature of $L_3$ in Fig.~\ref{fig:FigSRE}c (for details see Supplementary Information section~5).

We find that for data as in Fig.~\ref{fig:FigSRE}c the slopes of the emission lines are correlated by a set of sum rules
\begin{align}
  \Theta_{L3} &= \Theta_{L1} + \Theta_{L2}\label{eq:sumRule1}\\
  \Theta_{L4} &= 2\Theta_{L1} + \Theta_{L2}\label{eq:sumRule2}
\end{align}
Here $\Theta_{Ln}$ denotes the slope of emission line $L_n$ in Hertz per Tesla. The two-laser detuning frequencies for the pumping schemes in Fig.~\ref{fig:FigSchemes}a-d are related in the same way, which justifies the assignment of these four schemes to the emission lines $L_1$ through $L_4$, respectively. These schemes and equations directly yield the g-factor values $g_{g}$ and $g_{e}$ for the ground and excited state (Supplementary Information section~2).

We find that the g-factor values $g_{g}$ and $g_{e}$ strongly depend on $\phi$, that is, they are highly anisotropic. While this is in accordance with earlier observations for transition metal defects in SiC\cite{baur1997}, we did not find a comprehensive report on the underlying physical picture. In Supplementary Information section~7 we present a group-theoretical analysis that explains the anisotropy $g_{g} \approx 1.7$ for $\phi = \ang{0}$ and $g_{g} = 0$ for $\phi = \ang{90}$, and similar behavior for $g_{e}$ (which we also use to identify the orbital character of the ground and excited state). In this scenario the effective Land\'{e} g-factor\cite{abragam1970} is given by
\begin{equation}
\label{eq:geff}
g(\phi) = \sqrt{\left(g_{\parallel}\cos{\phi}\right)^2+\left(g_{\perp}\sin{\phi}\right)^2}
\end{equation}
where $g_{\parallel}$ represents the component of $g$ along the c-axis of the silicon carbide structure and $g_{\perp}$ the component in the basal plane. Figure~\ref{fig:FigGfactor} shows the ground and excited state effective g-factors extracted from our two-laser magnetospectroscopy experiments for 6H-SiC and 4H-SiC (additional experimental data can be found in the Supplementary Information). The solid lines represent fits to the equation (\ref{eq:geff}) for the effective g-factor. The resulting $g_{\parallel}$ and $g_{\perp}$ parameters are given in table \ref{tab:gfactors}.

The reason why diagonal transitions (in Fig.~\ref{fig:FigSchemes} panels a,c), and thus the $\Lambda$ and V scheme are allowed, lies in the different behavior of $g_e$ and $g_g$. When the magnetic field direction coincides with the internal quantization axis of the defect, the spin states in both the ground and excited state are given by the basis of the $S_z$ operator, where the $z$-axis is defined along the c-axis. This means that the spin-state overlap for vertical transitions, \textit{e.g.}~from $\ket{g_1}$ to $\ket{e_1}$, is unity. In such cases, diagonal transitions are forbidden as the overlap between \textit{e.g.}~$\ket{g_1}$ and $\ket{e_2}$ is zero. Tilting the magnetic field away from the internal quantization axis introduces mixing of the spin states. The amount of mixing depends on the g-factor, such that it differs for the ground and excited state. This results in a tunable non-zero overlap for all transitions, allowing all four schemes to be observed (as in Fig.~\ref{fig:FigSRE}b where $\phi = \ang{87}$). This reasoning also explains the suppression of all emission lines except $L_2$ in Fig.~\ref{fig:FigSRE}b, where the magnetic field is nearly along the c-axis. A detailed analysis of the relative peak heights in Fig.~\ref{fig:FigSRE}b-c compared to wave function overlap can be found in the Supplementary Information (section~4).

%%%%%%%%%%%%%%%%%%%%%%%%%%
%%%%%%%%% V) CPT
%%%%%%%%%%%%%%%%%%%%%%%%%%

\section{Coherent Population Trapping}

The $\Lambda$ driving scheme depicted in Fig.~\ref{fig:FigSchemes}a, where both ground states are coupled to a common excited state, is of particular interest. In such cases it is possible to achieve all-optical coherent population trapping (CPT)\cite{fleischhauer2005}, which is of great significance in quantum-optical operations that use ground-state spin coherence. This phenomenon occurs when two lasers address a $\Lambda$ system at exact two-photon resonance, \textit{i.e.}~when the two-laser detuning matches the ground-state splitting. The ground-state spin system is then driven towards a superposition state that approaches
$\ket{\Psi_{CPT}} \propto \Omega_2\ket{g_1} - \Omega_1\ket{g_2}$
for ideal spin coherence. Here $\Omega_n$ is the Rabi frequency for the driven transition from the $\ket{g_n}$ state to the common excited state. Since the system is now coherently trapped in the ground state, the photoluminescence decreases.

In order to study the occurrence of CPT, we focus on the two-laser PLE features that result from a $\Lambda$ scheme. A probe field with variable two-laser detuning relative to a fixed control laser was scanned across this line in frequency steps of 50~kHz, at $200~\mu\text{W}$. The control laser power was varied between $200~\mu\text{W}$ and $5~\text{mW}$. This indeed yields signatures of CPT, as presented in Fig.~\ref{fig:FigCPT}. A clear power dependence is visible: when the control beam power is increased, the depth of the CPT dip increases (and can fully develop at higher laser powers or by concentrating laser fields in SiC waveguides\cite{zwierThesis2016}). This observation of CPT confirms our earlier interpretation of lines $L_1$-$L_4$, in that it confirms that $L_1$ results from a $\Lambda$ scheme. It also strengthens the conclusion that this system is $S=1/2$, since otherwise optical spin-pumping into the additional (dark) energy levels of the ground state would be detrimental for the observation of CPT.

Using a standard model for CPT\cite{fleischhauer2005}, adapted to account for strong inhomogeneous broadening of the optical transitions\cite{zwierThesis2016} (see also Supplementary Information section~6) we extract an inhomogeneous spin dephasing time $T_2^*$ of $0.32\pm0.08$~$\mu$s and an optical lifetime of the excited state of $56\pm8$~ns. The optical lifetime is about a factor two longer than that of the nitrogen-vacancy defect in diamond\cite{dobrovitski2013,doherty2013}, indicating that the Mo defects can be applied as bright emitters (although we were not able to measure their quantum efficiency). The value of $T_2^*$ is relatively short but sufficient for applications based on CPT\cite{fleischhauer2005}.
Moreover, the EPR studies by Baur~\textit{et al.}\cite{baur1997} on various transition-metal impurities show that the inhomogeneity probably has a strong static contribution from an effect linked to the spread in mass for Mo isotopes in natural abundance (nearly absent for the mentioned vanadium case), compatible with elongating spin coherence via spin-echo techniques. In addition, their work showed that the hyperfine coupling to the impurity nuclear spin can be resolved. There is thus clearly a prospect for storage times in quantum memory applications that are considerably longer than $T_2^*$.

%%%%%%%%%%%%%%%%%%%%%%%%%%%%%%%%%%%%%%
%%%%%%%%% VI) FURTHER DISCUSSION
%%%%%%%%%%%%%%%%%%%%%%%%%%%%%%%%%%%%%%

\section{Further discussion}	

The anisotropic behavior of the g-factor that we observed for Mo was also observed for vanadium and titanium in the EPR studies by Baur~\textit{et al.}\cite{baur1997} (they observed $g_\parallel \approx 1.7$ and $g_\perp = 0$ for the ground state). In these cases the transition metal has a single electron in its 3d orbital and occupies the  hexagonal (\textit{h}) Si substitutional site. We show in Supplementary Information section~7 that the origin of this behavior can be traced back to a combination of a crystal field with $\rm C_{3v}$ symmetry and spin-orbit coupling for the specific case of an ion with one electron in its d-orbital.

The correspondence of this behavior with what we observe for the Mo impurity identifies that our materials have Mo impurities present as $\text{Mo}^{5+}$($4\text{d}^{1}$) systems residing on a hexagonal \textit{h} silicon substitutional site. In this case of a hexagonal (\textit{h}) substitutional site, the molybdenum is bonded in a tetrahedral geometry, sharing four electrons with its nearest neighbors. For $\text{Mo}^{5+}$($4\text{d}^{1}$) the defect is then in a singly ionized $+|e|$ charge state ($e$ denotes the elementary charge), due to the transfer of one electron to the p-type SiC host material.

An alternative scenario for our type of Mo impurities was recently proposed by Iv\'ady~\textit{et al.}\cite{ivady2011} They proposed, based on theoretical work\cite{ivady2011}, the existence of the asymmetric split-vacancy (ASV) defect in SiC. An ASV defect in SiC occurs when an impurity occupies the interstitial site formed by adjacent silicon and carbon vacancies. The local symmetry of this defect is a distorted octahedron with a threefold symmetry axis in which the strong g-factor anisotropy ($g_\perp = 0$) may also be present for the $S = 1/2$ state\cite{abragam1970}. Considering six shared electrons for this divacancy environment, the $\text{Mo}^{5+}$($4\text{d}^{1}$) Mo configuration occurs for the singly charged $-|e|$ state. For our observations this is a highly improbable scenario as compared to one based on the $+|e|$ state, given the p-type SiC host material used in our work. We thus conclude that this scenario by Iv\'ady~\textit{et al.} does not occur in our material. Interestingly, niobium defects have been shown to grow in this ASV configuration\cite{gallstrom2015}, indicating there indeed exist large varieties in the crystal symmetries involved with transition metal defects in SiC. This defect displays $S=1/2$ spin with several optical transitions between $892-897$~nm in 4H-SiC and $907-911$~nm in 6H-SiC\cite{gallstrom2015}.	

Another defect worth comparing to is the aforementioned chromium defect, studied by Koehl~\textit{et al.}\cite{koehl2017} Like Mo in SiC, the Cr defect is located at a silicon substitutional site, thus yielding a $3\text{d}^{2}$ configuration for this defect in its neutral charge state. The observed $S = 1$ spin state has a zero-field splitting parameter of 6.7~GHz\cite{koehl2017}. By employing optically detected magnetic resonance techniques they measured an inhomogeneous spin coherence time $T_2^*$ of $37$~ns\cite{koehl2017}, which is considerably shorter than observed for molybdenum in the present work. Regarding spin-qubit applications, the exceptionally low phonon-sideband emission of Cr seems favorable for optical interfacing. However, the optical lifetime for this Cr configuration ($146~\mu$s\cite{koehl2017}) is much longer than that of the Mo defect we studied, hampering its application as a bright emitter. It is clear that there is a wide variety in optical and spin properties throughout transition-metal impurities in SiC, which makes up a useful library for engineering quantum technologies with spin-active color centers.

%%%%%%%%%%%%%%%%%%%%%%%%%%%%%%%%%
%%%%%%%%%% VII) SUMMARY
%%%%%%%%%%%%%%%%%%%%%%%%%%%%%%%%%

\section{Summary and Outlook}
We have studied ensembles of molybdenum defect centers in 6H and 4H silicon carbide with 1.1521~eV and 1.1057~eV transition energies, respectively. The ground-state and excited-state spin of both defects was determined to be $S = 1/2$ with large g-factor anisotropy. Since this is allowed in hexagonal symmetry, but forbidden in cubic, we find this to be consistent with theoretical descriptions that predict that Mo resides at a hexagonal lattice site in 4H-SiC and 6H-SiC\cite{ivady2011,csore2016}, and
our p-type host environment strongly suggests that this occurs for Mo at a silicon substitutional site.
We used the measured insight in the $S=1/2$ spin Hamiltonians for tuning control schemes where two-laser driving addresses transitions of a $\Lambda$ system, and observed CPT for such cases. This demonstrates that the Mo defect and similar transition-metal impurities are promising for quantum information technology. In particular for the highly analogous vanadium color center, engineered to be in SiC material where it stays in its neutral $\text{V}^{4+}$($3\text{d}^{1}$) charge state, this holds promise for combining $S=1/2$ spin coherence with operation directly at telecom wavelengths.

%%%%%%%%%%%%%%%%%%%%%%%%%%%%%%%%%
%%%%%%%%%% METHODS
%%%%%%%%%%%%%%%%%%%%%%%%%%%%%%%%%

\vspace{1cm}
\noindent \textbf{Methods}\\

\begin{figure}[h!]
	\centering
	\includegraphics[width=10cm]{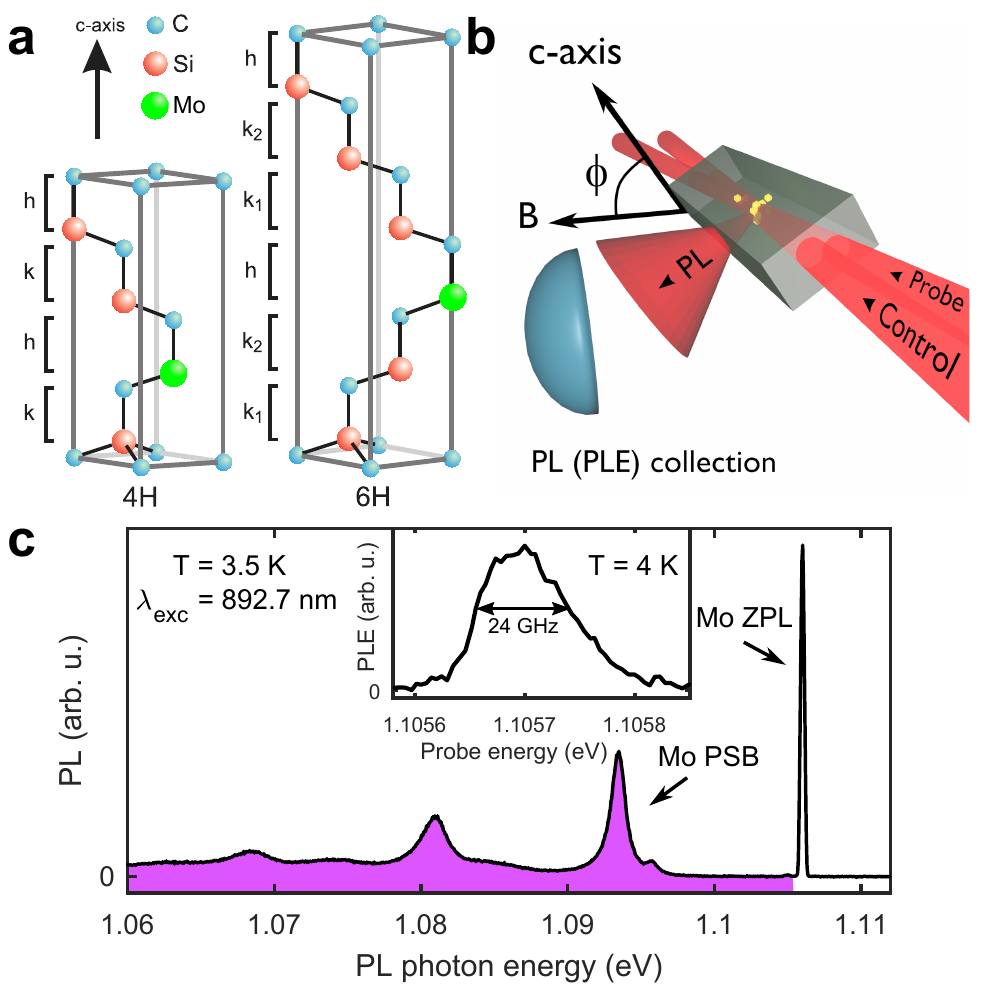}\\
	\caption{\textbf{Crystal structures of SiC, setup schematic and optical signatures of Mo in 6H-SiC.} \textbf{a}, Schematic illustration of the stacking of Si-C bilayers in the crystal structure of the 4H-SiC and 6H-SiC polytypes, which gives lattice sites with cubic and hexagonal local environment labeled by $k_{(1,2)}$ and $h$, respectively. Our work revisits the question whether Mo impurities are present as substitutional atoms (as depicted) or residing inside Si-C divacancies. The c-axis coincides with the growth direction. \textbf{b}, Schematic of SiC crystal in the setup. The crystal is placed in a cryostat with optical access. Laser excitation beams (control and probe for two-laser experiments) are incident on a side facet of the SiC crystal and propagate normal to the c-axis. Magnetic fields $\mathbf{B}$ are applied in a direction orthogonal to the optical axis and at angle $\phi$ with the c-axis. Photoluminescence (PL) is collected and detected out of another side facet of the SiC crystal. \textbf{c}, PL from Mo in 6H-SiC at 3.5~K and zero field, resulting from excitation with an 892.7~nm laser, with labels identifying the zero-phonon-line (ZPL, at 1.1057~eV) emission and phonon replicas (shaded and labeled as phonon sideband, PSB).
	The inset shows the ZPL as measured by photoluminescence excitation (PLE). Here, the excitation laser is scanned across the ZPL peak and emission from the PSB is used for detection.}
	\label{fig:FigPL}
\end{figure}

\noindent\textbf{Materials} The samples used in this study were $\sim$1~mm thick epilayers grown with chemical vapor deposition, and they were intentionally doped with Mo during growth. The PL signals showed that a relatively low concentration of tungsten was present due to unintentional doping from metal parts of the growth setup (three PL peaks near 1.00~eV, outside the range presented in Fig.~\ref{fig:FigPL}a). The concentration of various types of (di)vacancies was too low to be observed in the PL spectrum that was recorded. For more details see Ref.~\onlinecite{gallstrom2015thesis}.

\noindent\textbf{Cryostat} During all measurements, the sample was mounted in a helium flow cryostat with optical access through four windows and equipped with a superconducting magnet system.

\noindent\textbf{Photoluminescence (PL)} The PL spectrum of the 6H-SiC sample was measured by exciting the material with an 892.7~nm laser, and using a double monochromator equipped with infrared-sensitive photomultiplier.
For the 4H-SiC sample, we used a 514.5~nm excitation laser and an FTIR spectrometer.

\noindent\textbf{Photoluminescence Excitation (PLE)} The PLE spectrum was measured by exciting the defects using a CW diode laser tunable from 1050~nm to 1158~nm with linewidth below 50~kHz, stabilized within 1~MHz using feedback from a HighFinesse WS-7 wavelength meter. The polarization was linear along the sample c-axis. The laser spot diameter was $\sim$100~$\mu$m at the sample. The PL exiting the sample sideways was collected with a high-NA lens, and detected by a single-photon counter. The peaks in the PLE data were typically recorded at a rate of about 10~kcounts/s by the single-photon counter. We present PLE count rates in arb.~u.~since the photon collection efficiency was not well defined, and it varied with changing the angle $\phi$. For part of the settings we placed neutral density filters before the single-photon counter to keep it from saturating. The excitation laser was filtered from the PLE signals using a set of three 1082~nm (for the 4H-SiC case) or 1130~nm (for the 6H-SiC case) longpass interference filters. PLE was measured using an ID230 single-photon counter. Additionally, to counter charge state switching of the defects, a 770~nm re-pump beam from a tunable pulsed Ti:sapphire laser was focused at the same region in the sample. Laser powers as mentioned in the main text.

\noindent\textbf{Two-laser characterization} The PLE setup described above was modified by focusing a detuned laser beam to the sample, in addition to the present beams. The detuned laser field was generated by splitting off part of the stabilized diode laser beam. This secondary beam was coupled into a single-mode fiber and passed through an electro-optic phase modulator in which an RF signal (up to $\sim$5~GHz) modulated the phase. Several sidebands were created next to the fundamental laser frequency, the spacing of these sidebands was determined by the RF frequency. Next, a Fabry-P\'{e}rot interferometer was used to select one of the first-order sidebands (and it was locked to the selected mode). The resulting beam was focused on the same region in the sample as the original PLE beams (diode laser and re-pump) with similar spot size and polarization along the sample c-axis.  Laser powers were as mentioned in the main text. Small rotations of the c-axis with respect to the magnetic field were performed using a piezo-actuated goniometer with 7.2~degrees travel.

\noindent\textbf{Data processing} For all graphs with PLE data a background count rate is subtracted from each line, determined by the minimum value of the PLE in that line (far away from resonance features). After this a fixed vertical offset is added for clarity. For each graph, the scaling is identical for all lines within that graph.

%%%%%%%%%%%%%%%%%%%%%%%%%%%%%%%%%
%%%%%%%%%% DATA AVAILABILITY
%%%%%%%%%%%%%%%%%%%%%%%%%%%%%%%%%

\vspace{1cm}
\noindent \textbf{Data availability}\\
\noindent The data sets generated and analyzed during the current study are available from the corresponding author upon reasonable request.

%%%%%%%%%%%%%%%%%%%%%%%%%%%%%%%%%
%%%%%%%%%% ACKNOWLEDGEMENTS
%%%%%%%%%%%%%%%%%%%%%%%%%%%%%%%%%

\vspace{1cm}
\noindent \textbf{Acknowledgements}\\
We thank A.~Gali for discussions and M.~de~Roosz, J.~G.~Holstein, T.~J.~Schouten and H.~Adema for technical support.
Early discussions with Prof. Erik Janz\'{e}n leading to initiation of this study are gratefully acknowledged.
Financial support was provided by ERC Starting Grant 279931, the Zernike Institute BIS program, the Swedish Research Council grants VR~2016-04068 and VR~2016-05362, and the Carl-Trygger Stiftelse f\"{o}r Vetenskaplig Forskning grant CTS~15:339.

%%%%%%%%%%%%%%%%%%%%%%%%%%%%%%%%%%%%%%%%%%%%
%%%%%%%%%% COMPETING INTEREST STATEMENT
%%%%%%%%%%%%%%%%%%%%%%%%%%%%%%%%%%%%%%%%%%%%

\vspace{1cm}
\noindent \textbf{Competing interest}\\
\noindent The authors declare no competing financial or non-financial interests.

%%%%%%%%%%%%%%%%%%%%%%%%%%%%%%%%%%%%%%%%%%%%
%%%%%%%%%% AUTHOR CONTRIBUTIONS
%%%%%%%%%%%%%%%%%%%%%%%%%%%%%%%%%%%%%%%%%%%%

\vspace{1cm}
\noindent \textbf{Author Contributions}\\ The project was initiated by C.H.W., O.V.Z, I.G.I and N.T.S. SiC materials were grown and prepared by A.E. and B.M. Experiments were performed by T.B., G.J.J.L. and O.V.Z, except for the PL measurements which were done by A.G. and I.G.I. Data analysis was performed by T.B., G.J.J.L., C.G., O.V.Z., F.H., R.W.A.H. and C.H.W.  T.B., G.J.J.L. and C.H.W. had the lead on writing the paper, and T.B. and G.J.J.L. are co-first author. All authors read and commented on the manuscript.

%%%%%%%%%%%%%%%%%%%%%%%%%%%%%%%%%
%%%%%%%%%% REFERENCES
%%%%%%%%%%%%%%%%%%%%%%%%%%%%%%%%%

\vspace{1cm}
\noindent \textbf{REFERENCES FOR THE MAIN TEXT}

%%%%%%%%%%%%%%%%%%%%%%%%%%%%%%%%%%%%%%%%%%%%
%%%%%%%%%% FIGURES
%%%%%%%%%%%%%%%%%%%%%%%%%%%%%%%%%%%%%%%%%%%%

%--------------------------------------
%-------------- Figure 1 --------------
%--------------------------------------

\clearpage	

%--------------------------------------
%-------------- Figure 2 --------------
%--------------------------------------

\clearpage
\begin{figure}[h!]
	\centering
	\includegraphics[width=10cm]{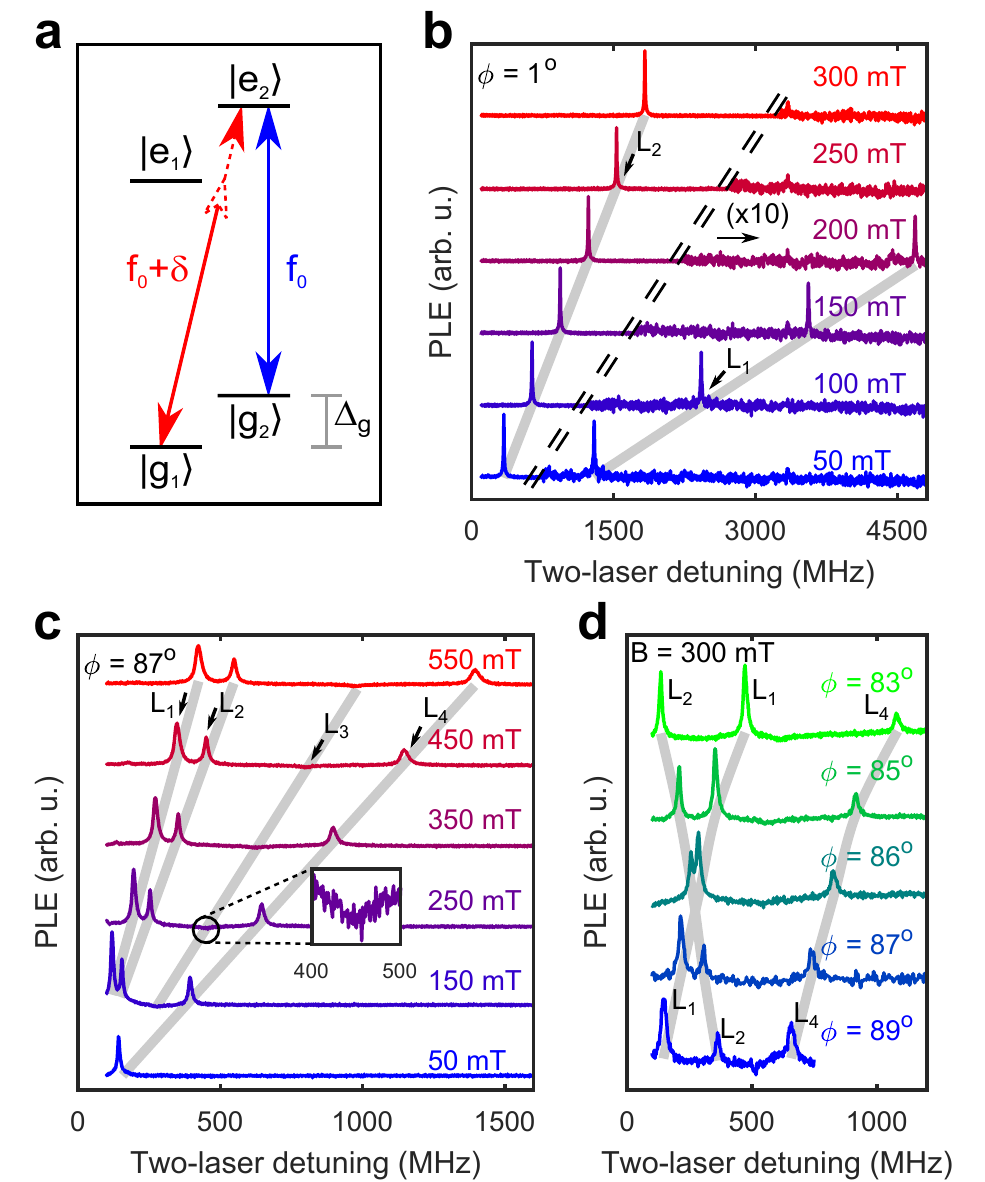}
	\caption{\textbf{Two-laser spectroscopy results for Mo in 6H-SiC}. \textbf{a}, Working principle of two-laser spectroscopy: one laser at frequency $f_0$ is resonant with the $\ket{g_2}$-$\ket{e_2}$ transition, the second laser is detuned from the first laser by $\delta$. If $\delta$ is such that the second laser becomes resonant with another transition (here sketched for $\ket{g_1}$-$\ket{e_2}$) the photoluminescence will increase since optical spin-pumping by the first laser is counteracted by the second and vice versa. \textbf{b-d}, Photoluminescence excitation (PLE) signals as a function of two-laser detuning at 4~K. \textbf{b}, Magnetic field dependence with field parallel to the c-axis ($\phi = \ang{1}$). For clarity, data in the plot have been magnified by a factor 10 right from the dashed line. Two peaks are visible, labeled $L_1$ and $L_2$ (the small peak at 3300~MHz is an artefact from the Fabry-P\'{e}rot interferometer in the setup). \textbf{c}, Magnetic field dependence with the field nearly perpendicular to the c-axis ($\phi = \ang{87}$). Three peaks and a dip (enlarged in the inset) are visible. These four features are labeled $L_1$ through $L_4$. The peak positions as a function of field in \textbf{b-c} coincide with straight lines through the origin (within 0.2\% error). \textbf{d}, Angle dependence of the PLE signal at 300~mT (angles accurate within $\ang{2}$). Peaks $L_1$ and $L_4$ move to the left with increasing angle, whereas $L_2$ moves to the right. The data in \textbf{b-d} are offset vertically for clarity.}
	\label{fig:FigSRE}
\end{figure}

%--------------------------------------
%-------------- Figure 3 --------------
%--------------------------------------

\clearpage
\begin{figure}[h!]
	\centering
	\includegraphics[width=13 cm]{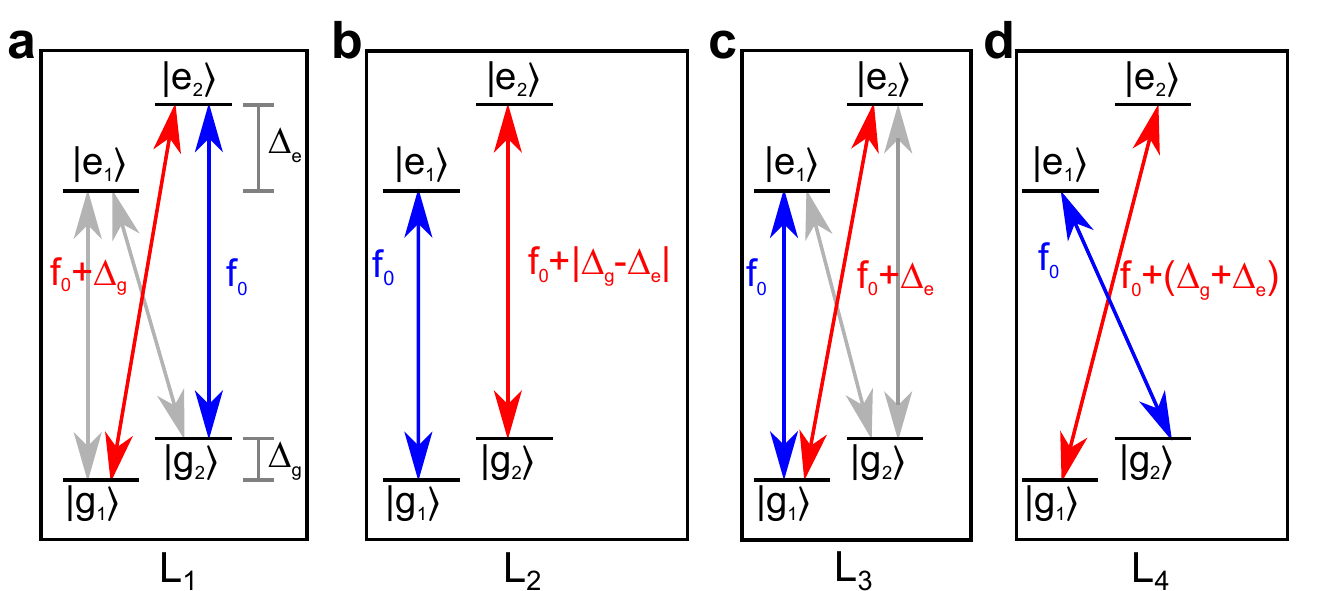}
	\caption{\textbf{Two-laser pumping schemes with optical transitions between \textit{S} = 1/2 ground and excited states}. \textbf{a}, $\Lambda$ scheme, responsible for $L_1$ emission feature: Two lasers are resonant with transitions from both ground states $\ket{g_1}$ (red arrow) and $\ket{g_2}$ (blue arrow) to a common excited state $\ket{e_2}$. This is achieved when the detuning equals the ground-state splitting $\Delta_g$. The gray arrows indicate a secondary $\Lambda$ scheme via $\ket{e_1}$ that is simultaneously driven in an ensemble when it has inhomogeneous values for its optical transition energies. \textbf{b}, $\Pi$ scheme, responsible for $L_2$ emission feature: Two lasers are resonant with both vertical transitions. This is achieved when the detuning equals the difference between the ground-state and excited-state splittings, $|\Delta_g-\Delta_e|$. \textbf{c}, V scheme, responsible for $L_3$ emission feature: Two lasers are resonant with transitions from a common ground state $\ket{g_1}$ to both excited states $\ket{e_1}$ (blue arrow) and $\ket{e_2}$ (red arrow). This is achieved when the laser detuning equals the excited state splitting $\Delta_e$. The gray arrows indicate a secondary V scheme that is simultaneously driven when the optical transition energies are inhomogeneously broadened. \textbf{d}, X scheme, responsible for the $L_4$ emission feature: Two lasers are resonant with the diagonal transitions in the scheme. This is achieved when the detuning is equal to the sum of the ground-state and the excited-state splittings, $(\Delta_g + \Delta_e)$.}
	\label{fig:FigSchemes}
\end{figure}

%--------------------------------------
%-------------- Figure 4 --------------
%--------------------------------------

\clearpage
\begin{figure}[h!]
	\centering
	\includegraphics[width=10cm]{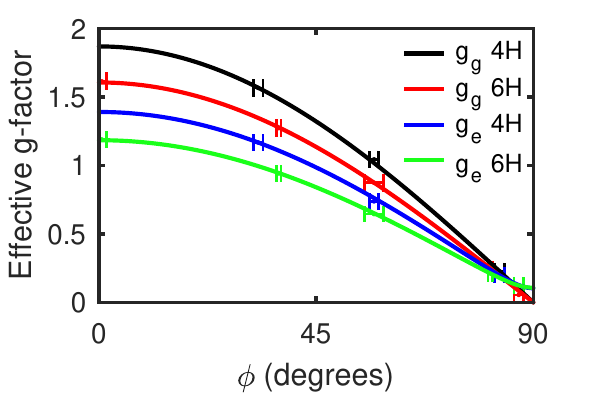} % width original picture = 6cm
	\caption{\textbf{Effective g-factors for the spin of Mo impurities in SiC.} Angular dependence of the g-factor for the $S = 1/2$ ground ($g_g$) and excited states ($g_e$) of the Mo impurity in 4H-SiC and 6H-SiC. The solid lines indicate fits of equation (\ref{eq:geff}) to the data points extracted from two-laser magneto-spectroscopy measurements as in Fig.~\ref{fig:FigSRE}b,c.}
		\label{fig:FigGfactor}
\end{figure}

%--------------------------------------
%-------------- Figure 5 --------------
%--------------------------------------

\clearpage
\begin{figure}[h!]
	\centering
	\includegraphics[width=10cm]{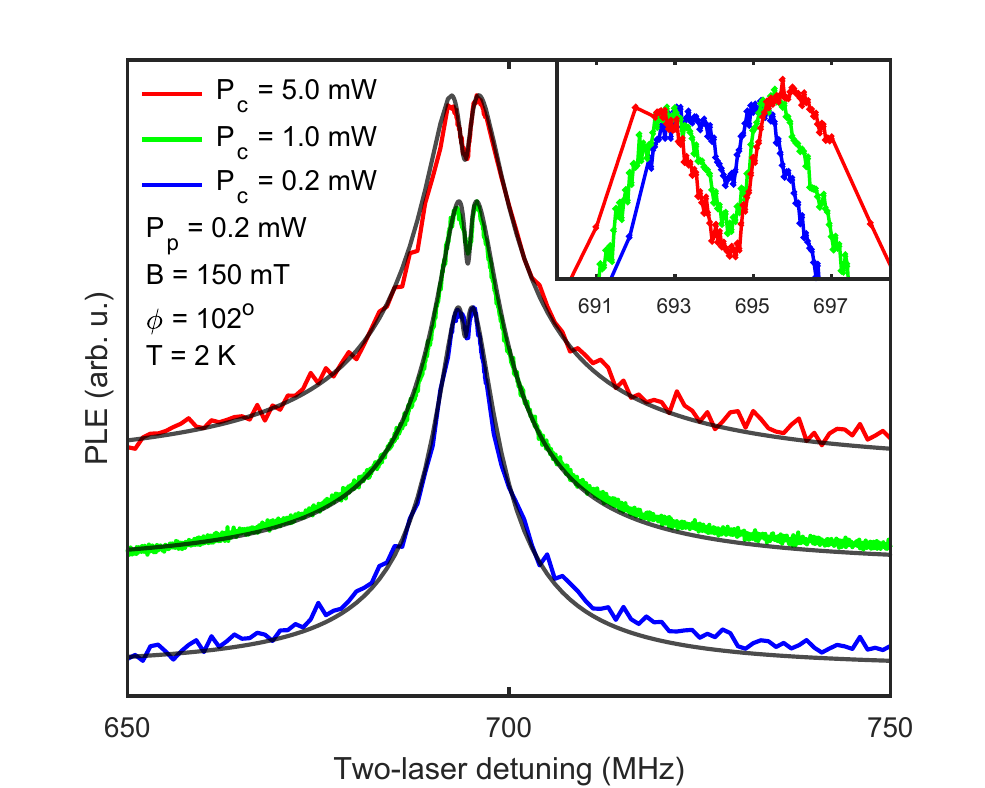}
    \caption{\textbf{Signatures of coherent population trapping of Mo spin states in 6H-SiC}. Two-laser spectroscopy of the $L_1$ peak in the PLE signals reveals a dipped structure in the peak at several combinations of probe-beam and control-beam power. This results from coherent population trapping (CPT) upon $\Lambda$-scheme driving. Temperature, magnetic field orientation and magnitude, and laser powers, were as labeled. The data are offset vertically for clarity. Solid lines are fits of a theoretical model of CPT (see main text). The inset shows the normalized CPT feature depths.}
		\label{fig:FigCPT}
\end{figure}

%%%%%%%%%%%%%%%%%%%%%%%%%%%%%%%%%%%%%%%%%%%%%
%%%%%%%%%%% TABLES
%%%%%%%%%%%%%%%%%%%%%%%%%%%%%%%%%%%%%%%%%%%%%

%--------------------------
%------- Table 1 ----------
%--------------------------

\begin{table}
	\caption{\textbf{Components of the g-factors for the spin of Mo impurities in SiC}}
	\label{tab:gfactors}
	\begin{center}
		\begin{tabular}{ l@{\hskip 10mm} l@{\hskip 10mm} l }
			\hline
			& $g_\parallel$ & $g_\perp$\\
			\hline
			4H-SiC\\
			ground state & $1.87\pm0.2$ & $0.04\pm0.04$\\
			excited state & $1.39\pm0.2$ & $0.10\pm0.02$\\
			6H-SiC\\
			ground state & $1.61\pm0.02$ & $0.000\pm0.004$\\
			excited state & $1.20\pm0.02$ & $0.11\pm0.02$\\
			\hline
		\end{tabular}
	\end{center}
\end{table}

\clearpage
%%%%%%%%%%%%%%%%%%%%%%%%%%%%%%%%%%%%%%%%%%%%%
%%%%%%%%%%% Supplemental Information
%%%%%%%%%%%%%%%%%%%%%%%%%%%%%%%%%%%%%%%%%%%%%

\setcounter{table}{0}
\setcounter{figure}{0}
\setcounter{section}{0}
\setcounter{page}{1}
\setcounter{tocdepth}{1}

\renewcommand{\thesection}{\arabic{section}}
\renewcommand{\thefigure}{S\arabic{figure}}
\renewcommand{\thetable}{S\arabic{table}}

\makeatletter
\renewcommand\@biblabel[1]{#1.}
\makeatother

%\renewcommand{\refname}{xx} % Was not needed since there was a manual heading

%FOR SDFT Section
\newcommand{\be}[1]{\begin{eqnarray}  {\label{#1}}}
	\newcommand{\ee}{\end{eqnarray}}

\begin{center}
	\textbf{{\LARGE Supplementary Information}}
	
	for
	
	\textbf{{\Large Identification and tunable optical coherent control of transition-metal spins in silicon carbide}}
	
	by
	
	Tom~Bosma *, Gerrit~J.~J.~Lof *,Carmem~M.~Gilardoni, Olger~V.~Zwier, Freddie~Hendriks, Bj\"orn Magnusson, Alexandre Ellison, Andreas G\"allstr\"om, Ivan G. Ivanov, N.~T.~Son, Remco~W.~A.~Havenith, and Caspar~H.~van~der~Wal
	
\end{center}

\vspace{4cm}

\noindent \textbf{TABLE OF CONTENTS}
%\tableofcontents

\begin{figure}[h!]
	\includegraphics[width=\columnwidth]{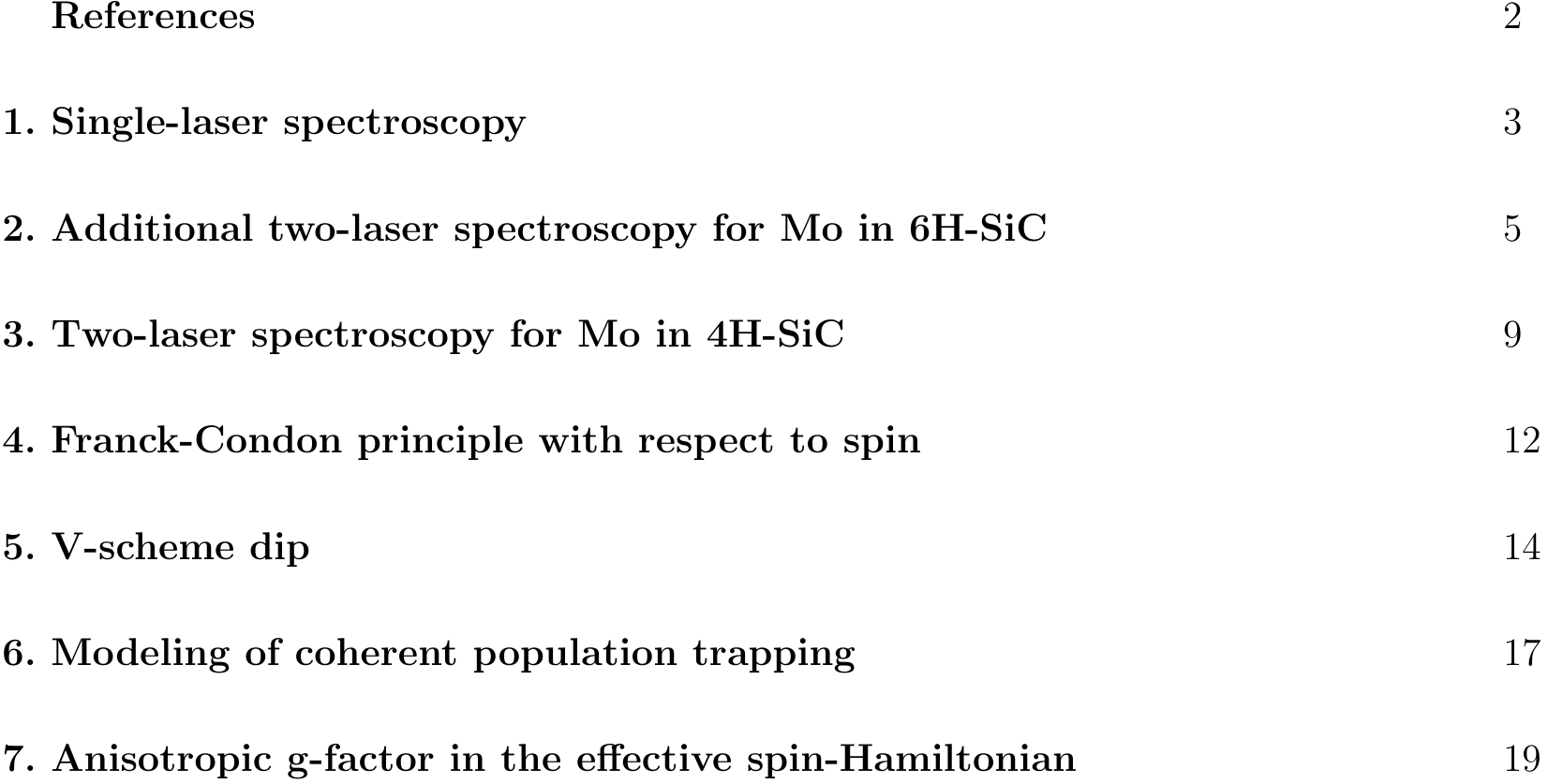} %%% Needs an update
\end{figure}

%%%%%%%%%%%%%%%% END ARXIV TOP PART

\newpage

%\noindent \textbf{References and Notes}

\noindent \textbf{REFERENCES FOR THE SUPPLEMENTARY INFORMATION}

\newpage

\section{Single-laser spectroscopy}

\begin{figure}[h!]
	\centering
	\includegraphics[width=11.6cm]{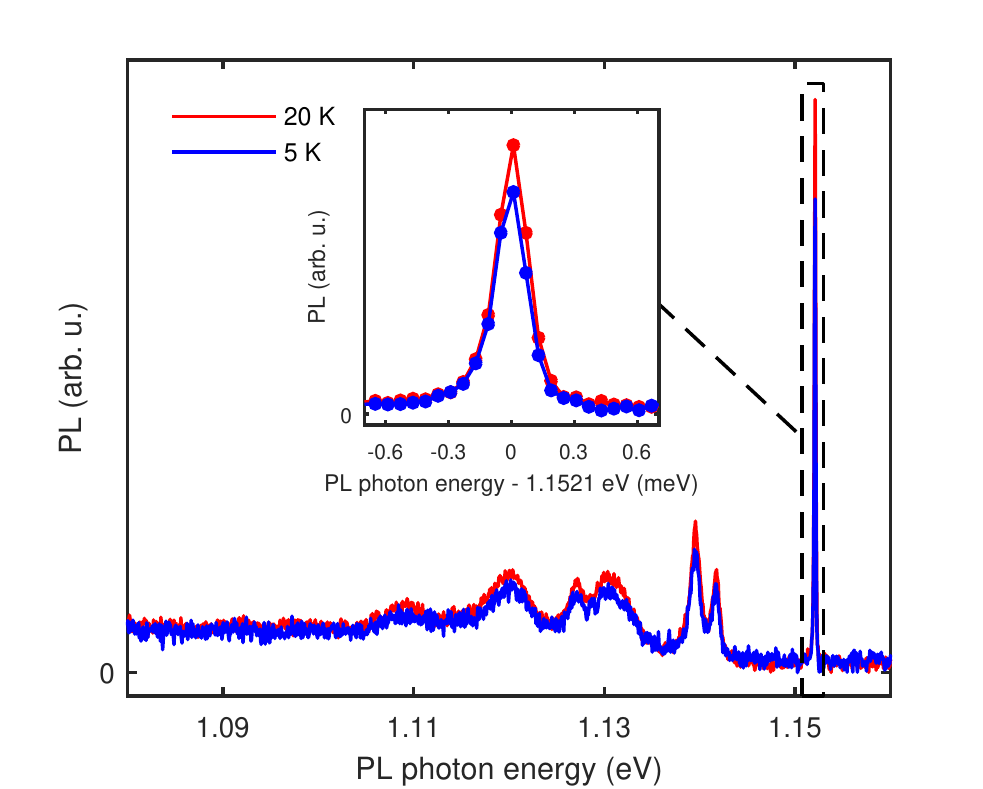}
	\caption{\textbf{Temperature dependence of Mo PL spectrum in 4H-SiC.} %Molybdenum impurity photoluminescence.
		PL from excitation with a 514.5~nm laser, for 5 and 20~K sample temperatures. The dashed box marks the ZPL at 1.1521~eV. The inset gives a magnified view of the ZPL. The broader peaks at lower photon energies are phonon replicas of the ZPL. %There is almost no dependence on temperature in both ZPL and replicas.
		%Data courtesy of A. G{\"a}llstr{\"o}m, Department of Physics, Chemistry and Biology, Link{\"o}ping University, Sweden
	}
	\label{fig:FigPL4H}
\end{figure}

Figure~\ref{fig:FigPL4H} shows the photoluminescence (PL) emission spectrum of the 4H-SiC sample at 5 and 20~K, characterized using a 514.5~nm
excitation laser. The Mo zero-phonon line (ZPL) at 1.1521~eV is marked by a dashed box and shown enlarged in the inset. The broader peaks at lower energies are phonon replicas of the ZPL. There is almost no dependence on temperature for both the ZPL and the phonon replicas.

%%%%%%%%%%%%%%%%%%%%%%%%%%%%%%%%%%%%%%%
%%%%%%%%%%%%%%% Figure S1 %%%%%%%%%%%%%
%%%%%%%%%%%%%%%%%%%%%%%%%%%%%%%%%%%%%%%

% Temperature dependence of PL in 4H and 6H-SiC.

\begin{figure}[h!]
	\centering
	\includegraphics[width=10cm]{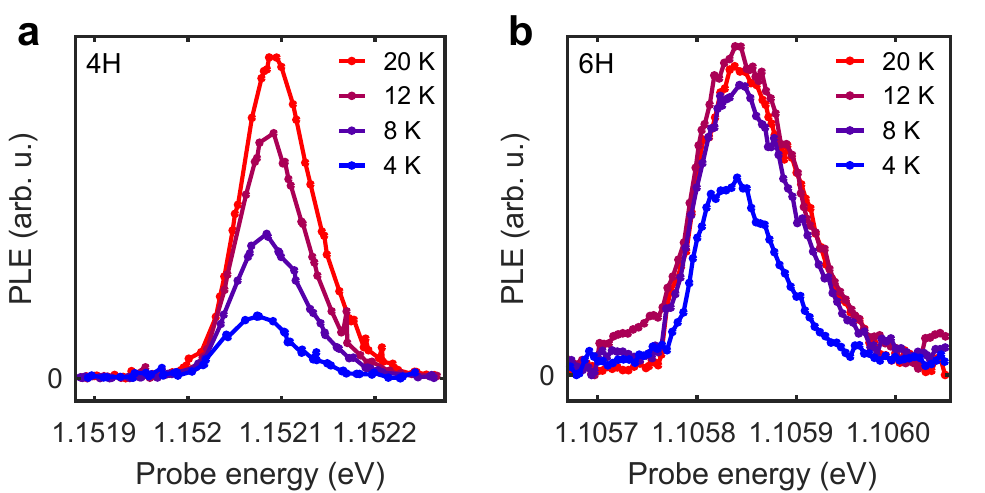}
	\caption{\textbf{Temperature dependence of the PLE signals from the Mo ZPL in 4H-SiC and 6H-SiC.}
		PLE signals from scanning a single CW narrow-linewidth laser across the ZPL photon-energy range. The temperature was varied between 4 and 20~K. The ZPL for Mo in (\textbf{a}) 4H-SiC is at 1.1521~eV, and for Mo in (\textbf{b}) 6H-SiC at 1.1057~eV. %The peak position is shifted with 0.2~meV compared to Fig.~\ref{fig:FigPL4H}, which is within the uncertainty of \textbf{[waiting for reply collaborator]}. Experiments with the laser resonant with the ZPL were performed within 0.1~meV from the ZPL peak.
	}
	\label{fig:FigPLTemp}
\end{figure}

Figures~\ref{fig:FigPLTemp}a,b show results of PLE measurements of the ZPL for Mo in 4H-SiC at 1.1521~eV and 6H-SiC at 1.1057~eV, and the temperature dependence of these PLE signals.
%which is in close agreement with literature\cite{gallstrom2009}.
%NB: these values are slightly inaccurate since the WLmeter has not been calibrated for some time. However, this does not affect the relative accuracy of all measurements.
When the temperature is decreased, the width of the ZPL stays roughly the same, but its height drops significantly. Combined with the near-independence on temperature of the emission spectrum in Fig.~\ref{fig:FigPL4H}, this is an indication for optical spin pumping for Mo-impurity states at lower temperatures, where a single resonant laser pumps then the system into long-lived off-resonant spin states.

%%%%%%%%%%%%%%%%%%%%%%%%%%%%%%%%%%%%%%%
%%%%%%%%%%%%%%% Figure S2 %%%%%%%%%%%%%
%%%%%%%%%%%%%%%%%%%%%%%%%%%%%%%%%%%%%%%

% Temperature dependence of PLE in 4H and 6H-SiC.
\clearpage

\section{Additional two-laser spectroscopy for ${\rm {\bf Mo}}$ in 6H-${\rm {\bf SiC}}$}\label{sec:6HSiCSI}
\begin{figure}[h!]
	\centering
	\includegraphics[width=17cm]{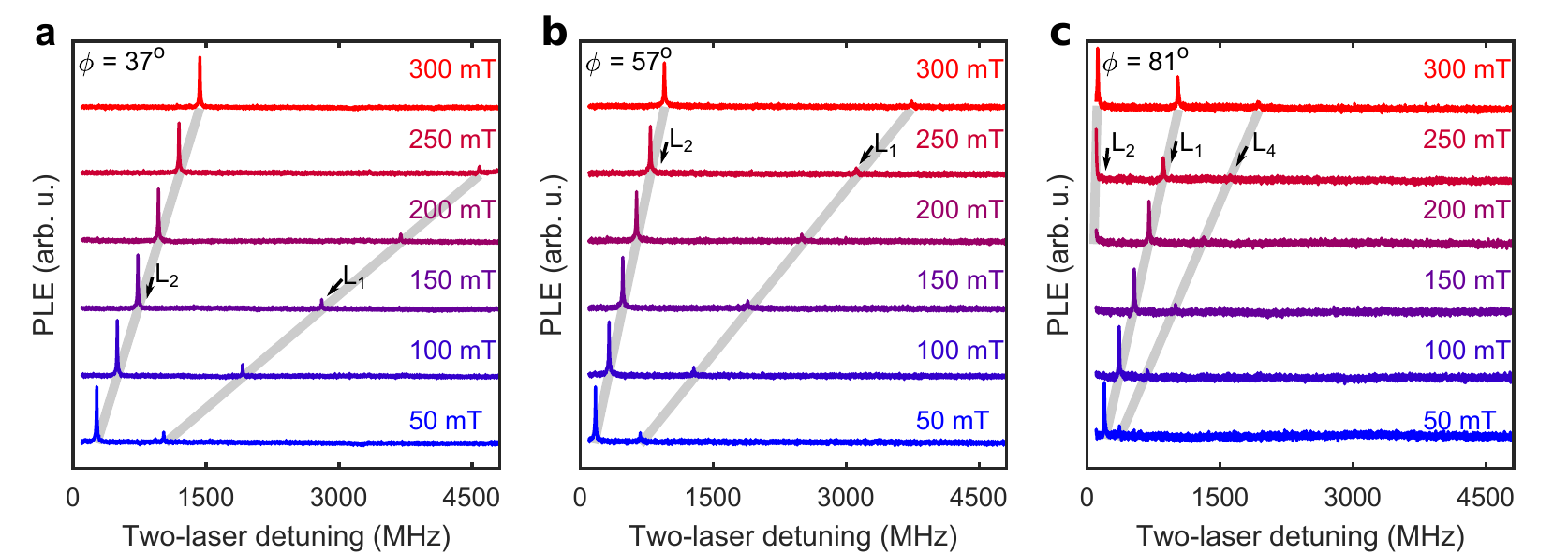}
	\caption{\textbf{Magneto-spectroscopy of two-laser spin signatures in PLE from Mo in 6H-SiC.} Magnetic field dependence of the PLE signal as a function of two-laser detuning, for angles $\phi$ between the magnetic field and c-axis set to $\phi = \ang{37}$ (\textbf{a}), $\phi = \ang{57}$  (\textbf{b}) and $\phi = \ang{81}$ (\textbf{c}). Results for the temperature at 4~K. The labeling of the emission lines ($L_1$ - $L_4$) is consistent with Fig.~\ref{fig:FigSRE}. The data are offset vertically for clarity.}
	\label{fig:FigMoreSRE}
\end{figure}

%\paragraph{Mo in 6H-SiC}
\paragraph{Angle dependence.} In addition to Fig.~\ref{fig:FigSRE}b,c in the main text, we also measured the magnetic field dependence of the spin related emission signatures at intermediate angles $\phi$. Figure~\ref{fig:FigMoreSRE} shows this dependence for $\phi = \ang{37}$, $\ang{57}$ and $\ang{81}$. The spectroscopic position of emission lines $L_n$ show a linear dependence on magnetic field, with slopes $\Theta_{Ln}$ (in Hertz per Tesla) that decrease as $\phi$ increases. The effective g-factors in Fig.~\ref{fig:FigGfactor} are acquired from the emission lines by relating their slopes to the Zeeman splittings in the ground and excited state. Using the four pumping schemes depicted in Fig.~\ref{fig:FigSchemes} in the main text, we derive
\begin{align}
	\Theta_{L1}  	    &=  \frac{\mu_B}{h} g_g \label{eq:L1}\\
	\Theta_{L2}		&= \frac{\mu_B}{h} |g_e - g_g| \label{eq:L2}\\
	\Theta_{L3}		&= \frac{\mu_B}{h} g_e \label{eq:L3}\\
	\Theta_{L4}		&= \frac{\mu_B}{h}  \left(g_e + g_g\right) \label{eq:L4}
\end{align} where $h$ is Planck's constant, $\mu_B$ the Bohr magneton and $g_{g(e)}$ the ground (excited) state g-factor.

%The effective g-factors are acquired using the same reasoning as in the {\color{red}main text}, they are shown in Fig.~\ref{fig:FigGfactor} and used to fit the effective anisotropic g-factor from equation~\ref{eq:geff}.

%%%%%%%%%%%%%%%%%%%%%%%%%%%%%%%%%%%%%%%
%%%%%%%%%%%%%%% Figure S3 %%%%%%%%%%%%%
%%%%%%%%%%%%%%%%%%%%%%%%%%%%%%%%%%%%%%%

% More SRE resutls for 6H-SiC

\paragraph{Temperature and photon-energy dependence.}

\begin{figure}[h!]
	\centering
	\includegraphics[width=12cm]{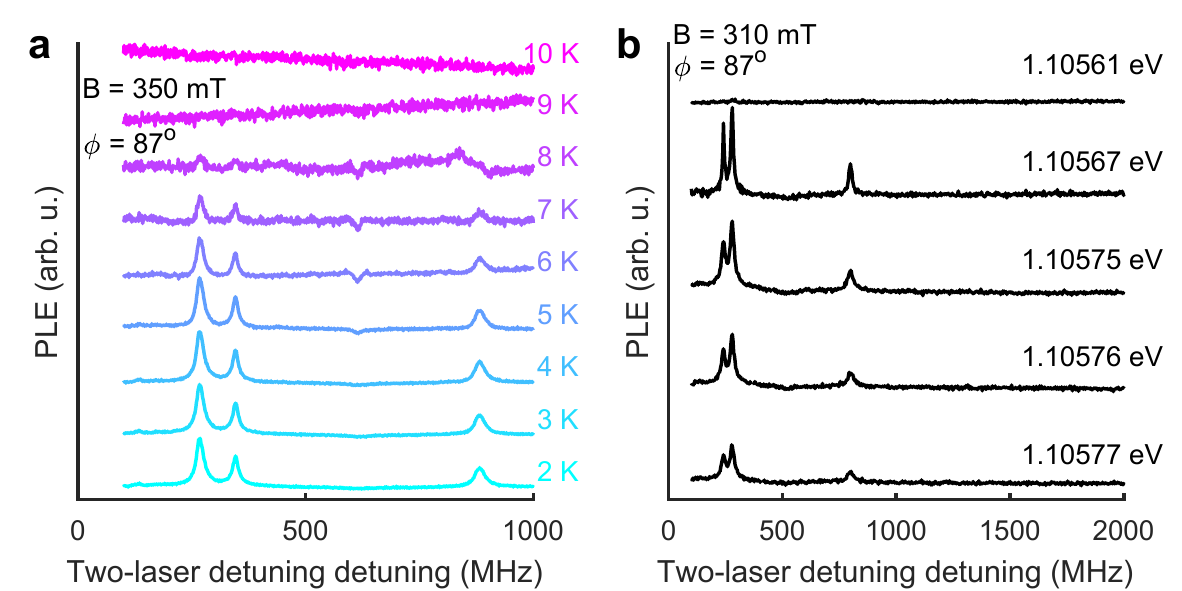}
	\caption{\textbf{Temperature and photon-energy dependence of two-laser emission features in 6H-SiC.} Temperature (\textbf{a}) and laser photon-energy (\textbf{b}) dependence of the PLE signal as a function of two-laser detuning. Results are offset vertically for clarity.}
	\label{fig:FigSREWLTemp}
\end{figure}

We also measured the dependence of the two-laser PLE signal on temperature, see Fig.~\ref{fig:FigSREWLTemp}a. The PLE features disappear above 8~K in a much broader PLE background that starts to emit because of more rapid thermal spin mixing in the ground state. Other Mo systems in the ensemble for which the two-laser resonance condition is not met then also start emitting due to single-laser excitation (see also Fig.~\ref{fig:FigPLTemp}), since the optical pumping into an off-resonant state becomes shorter lived. Notably, the linewidths of the peaks in Fig.~\ref{fig:FigSREWLTemp}a do not change in the range 2~K to 8~K, indicating that the temperature does not affect the optical lifetime in this range.
Above 8~K the temperature was a bit unstable during the measurements, which causes the drifting in the single-laser PLE contribution to this signal. Interestingly, the dip ($L_3$) is most pronounced at 6~K, since there are several competing processing responsible for this dip (see section~\ref{sec:SI-dip}).

Additionally, we measured how the two-laser PLE occurred throughout the inhomogenously broadened ensemble, by varying the photon energy of the control laser and sweeping the two-laser detuning for each case (Fig.~\ref{fig:FigSREWLTemp}b). For all photon energies the peaks are at the same position, indicating that all Mo atoms in the ensemble behave similarly. At 1.10561~eV the control laser is too far detuned from the ZPL to yield any two-laser PLE signal.

\newpage 

\begin{figure}[h!]
	\centering
	\includegraphics[width=10cm]{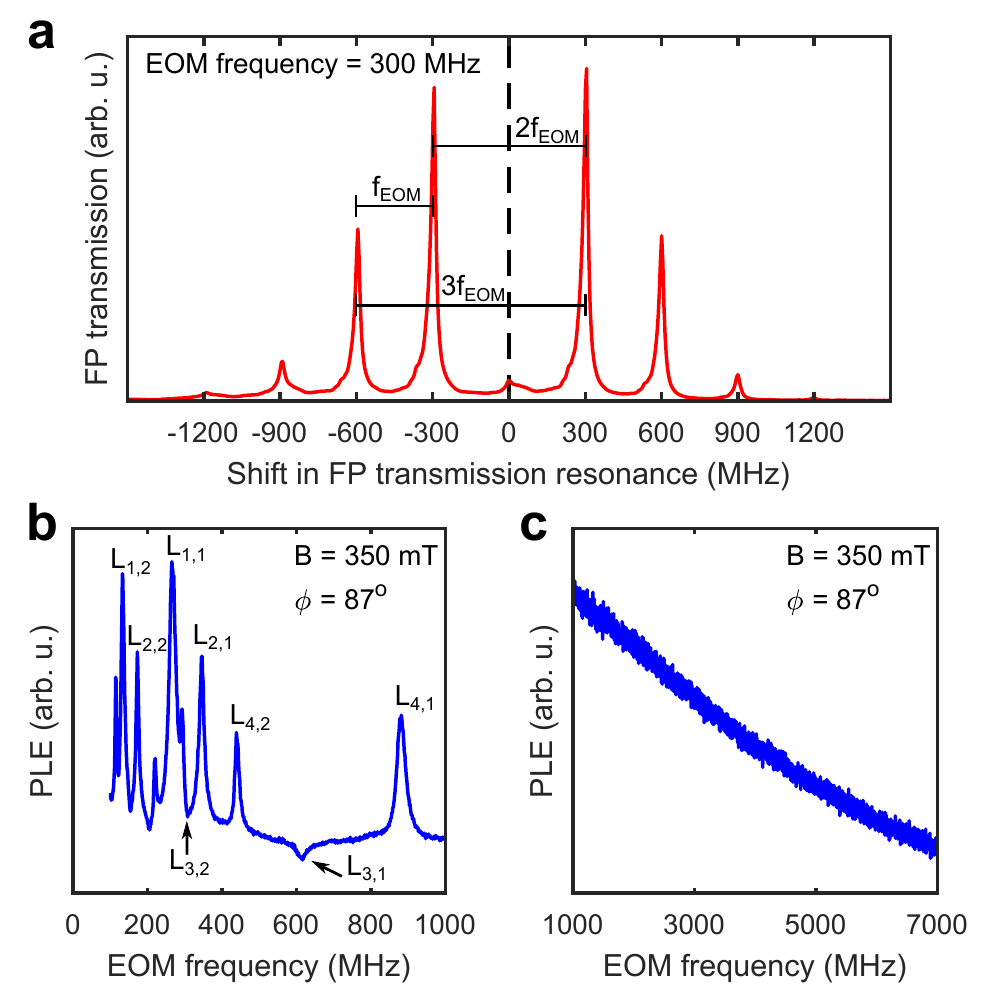}
	\caption{\textbf{Two-laser spin signatures of Mo in 6H-SiC at large detuning.} \textbf{a}, Transmission scan of the Fabry-P\'{e}rot resonator, characterizing which optical frequencies are present in the beam after passing through the electro-optical modulator (EOM). The first-order sidebands at $\pm300~\text{MHz}$ have the highest intensity, whereas the fundamental laser frequency is suppressed (but not fully removed) by the EOM. Relevant sideband spacings are indicated. \textbf{b}, Spin signatures at low two-laser detuning. PLE is increased when two sidebands are appropriately detuned from each other. Emission features similar to those in Fig.~\ref{fig:FigSRE}c of the main text are visible, and labeled $L_{n,m}$ (see main text of this section). \textbf{c}, The PLE signal from two-laser spectroscopy at larger detuning. No peaked features from single, double or triple sideband spacings are visible.}
	\label{fig:FigLargeDetuning}
\end{figure}

\paragraph{Two-laser spectroscopy for the 5-21~GHz detuning range.} In order to check for a possible presence of spin-related emission features at detunings larger than 5~GHz (checking for a possible zero-field splitting), we modified the setup such that we could control two-laser detunings up to 21~GHz. The electro-optical phase modulator (EOM) we used for generating the detuned laser field could generate first-order sidebands up to 7~GHz. In order to check for two-laser spectroscopy emission features at larger detunings, we removed the Fabry-P\'{e}rot (FP) resonator that had the role of filtering out a single sideband. Now, all sidebands (on the same optical axis) were focused onto the sample with 2~mW total laser power. Apart from the re-pump beam, no additional laser was focused onto the sample in this experiment. In this way, the Mo defects could interact with several combinations of sidebands. Figure~\ref{fig:FigLargeDetuning}a shows the spectral content of this beam (here characterized by still using the FP resonator). The first and second order sidebands at negative and positive detuning take a significant portion of the total optical power. Hence, pairs of sidebands spaced by single, double or triple frequency intervals (EOM frequency $f_\text{EOM}$) now perform two-laser spectroscopy on the Mo defects. The relevant sideband spacings are indicated in Fig.~\ref{fig:FigLargeDetuning}a.

Figure~\ref{fig:FigLargeDetuning}b presents results of these measurements, showing various peaks that we identified and label as $L_{n,m}$. Here $n$ is identifying the peak as a line $L_n$ as in the main text, while the label $m$ identifies it as a spectroscopic response for two-laser detuning at $m \cdot f_\text{EOM}$ (that is, $m=1$ is for first-order EOM sideband spacing, \textit{etc.}). Note that second-order manifestations of the known peaks $L_1$-$L_4$ (from double sideband spacings, labeled as $L_{n,2}$) are now visible at $\frac{1}{2}f_\text{EOM}$, and third-order response of the known $L_1$-$L_4$ occurs at $\frac{1}{3}f_\text{EOM}$ (but for preserving clarity these have not been labeled in Fig.~\ref{fig:FigLargeDetuning}b).

Figure~\ref{fig:FigLargeDetuning}c depicts a continuation of this experiment with $f_\text{EOM}$ up to 7~GHz with the same resolution as Fig.~\ref{fig:FigLargeDetuning}b. No new peaks are observed. Considering that third-order peaks were clearly visible before, we conclude that no additional two-laser emission features exist up to 21~GHz.

%%%%%%%%%%%%%%%%%%%%%%%%%%%%%%%%%%%%%%%
%%%%%%%%%%%%%%% Figure S4 %%%%%%%%%%%%%
%%%%%%%%%%%%%%%%%%%%%%%%%%%%%%%%%%%%%%%

\clearpage

\section{Two-laser spectroscopy for ${\rm {\bf Mo}}$ in 4H-${\rm {\bf SiC}}$}

\begin{figure}[h!]
	\centering
	\includegraphics[width=13cm]{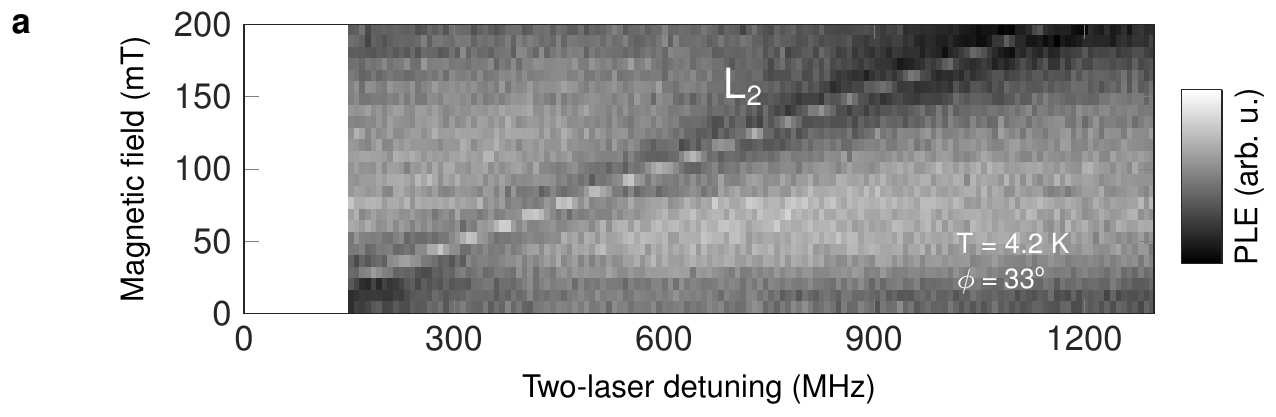}
	\includegraphics[width=13cm]{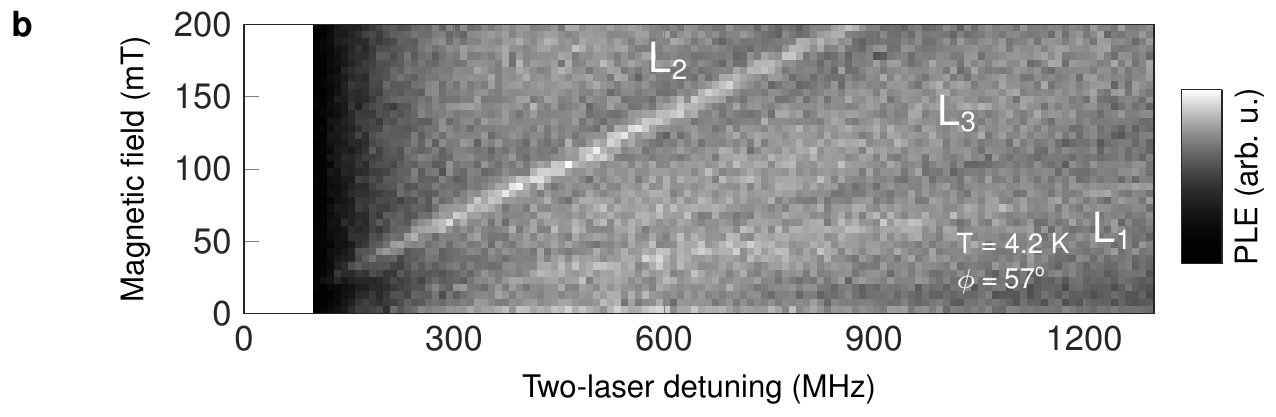}
	\includegraphics[width=13cm]{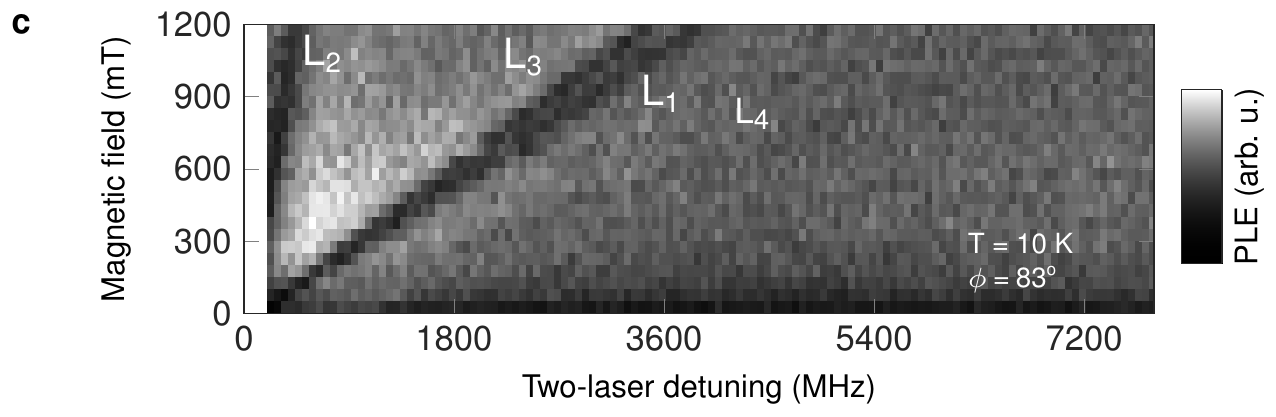}
	\caption{\textbf{Two-laser spin signatures of Mo in 4H-SiC.} PLE signal as a function of two-laser detuning and magnetic field strength, for various angles $\phi$ between the magnetic field and c-axis. \textbf{a}, Measurement at 4.2~K, with $\phi = \ang{33}$. A single emission line (peak) is visible, labeled $L_2$. \textbf{b}, Measurement at 4.2~K, with $\phi = \ang{57}$. Three emission lines are visible, labeled $L_1$, $L_2$ (peaks), and $L_3$ (dip). \textbf{c}, Measurement at 10~K, with $\phi = \ang{83}$. Four emission lines are visible, labeled $L_1$ through $L_4$ (all dips). Note that the measurement range of \textbf{c} is six time as large as \textbf{a} and \textbf{b}, but the plot aspect ratio is the same. The labeling is consistent with the main text.  A gray-scale plot has been used for optimal contrast.}
	\label{fig:FigSRE4H}
\end{figure}

%\paragraph{Mo in 4H-SiC.}
We also studied the spin-related fine structure of Mo defects in 4H-SiC. Our 4H-SiC sample suffered from large background absorption, which drastically lowered the signal-to-noise ratio. We relate this absorption to a larger impurity content (of unknown character, but giving broad-band absorption) in our 4H-SiC material as compared to our 6H-SiC material. Therefore, the lasers were incident on a corner of the sample, so as to minimize the decay of the emitted PL. We present the results in gray-scale plots in Fig.~\ref{fig:FigSRE4H} for optimized contrast. The figure shows the magnetic field and two-laser detuning dependence of the PLE.

Analogous to Fig.~\ref{fig:FigSRE} for 6H-SiC in the main text, the spectroscopic position features appear as straight lines that emerge from zero detuning, indicating the absence of a zero-field splitting. When the magnetic field is nearly perpendicular to the c-axis (Fig.~\ref{fig:FigSRE4H}c), four lines are visible. This is consistent with an $S = 1/2$ ground and excited state.

%%%%%%%%%%%%%%%%%%%%%%%%%%%%%%%%%%%%%%%
%%%%%%%%%%%%%%% Figure S5 %%%%%%%%%%%%%
%%%%%%%%%%%%%%%%%%%%%%%%%%%%%%%%%%%%%%%

% More SRE resutls for 4H-SiC

\begin{figure}[h!]
	\centering
	\includegraphics[width=6cm]{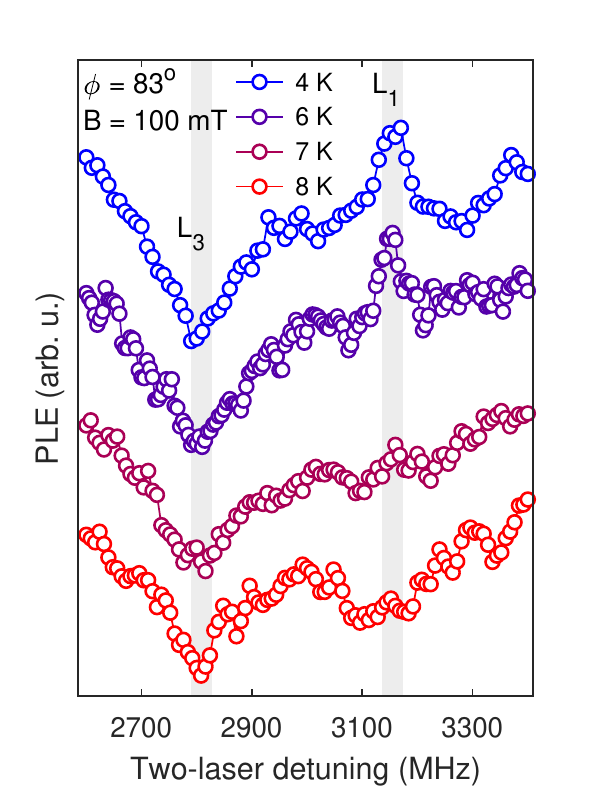}
	\caption{\textbf{Temperature dependence of PLE spin signatures from Mo in 4H-SiC.} PLE signal as a function of two-laser detuning and temperature with magnetic field at $\phi=\ang{83}$ from the sample c-axis at 100~mT. As the temperature increases, the signal from $L_1$ changes from a peak to a broad dip, while $L_3$ remains a dip. The labeling is consistent with the main text.}\label{fig:FigSRE4HTemp}
\end{figure}

The data from Fig.~\ref{fig:FigSRE4H}c was measured at 10~K, whereas Fig.~\ref{fig:FigSRE4H}a,b was at 4.2~K. At 10~K, all emission lines become dips, while for 6H-SiC only the V system shows a dip. The temperature dependence of $L_3$ and $L_1$ is shown in Fig.~\ref{fig:FigSRE4HTemp} for the same configuration as in Fig.~\ref{fig:FigSRE4H}c ($\phi = \ang{83}$). At low temperatures $L_1$ shows a peak and $L_3$ shows a dip. Upon increasing the temperature, both features become dips. This phenomenon was only observed for Mo in 4H-SiC, it could not be seen in 6H-SiC. We therefore conclude that this probably arises from effects where Mo absorption and emission is influenced by the large background absorption in the 4H-SiC material.

%%%%%%%%%%%%%%%%%%%%%%%%%%%%%%%%%%%%%%%
%%%%%%%%%%%%%%% Figure S6 %%%%%%%%%%%%%
%%%%%%%%%%%%%%%%%%%%%%%%%%%%%%%%%%%%%%%
% Analysis of wave function overlap: Probability of exciting in two-laser schemes.
% NB: Since g_perp is zero for ground state, there is no mixing in the ground state. The eigenstates are always in the basis is S_z.

% More SRE resutls for 4H-SiC

The labels in Fig.~\ref{fig:FigSRE4H} are assigned based on the sum rules from equation (\ref{eq:sumRule1}) and (\ref{eq:sumRule2}) (main text), which indeed also hold for the observed emission lines observed here. Like in the main text, $L_1$ through $L_4$ indicate $\Lambda$, $\Pi$, V and X two-laser pumping schemes, respectively. The $L_1$ and $L_3$ labels  are interchangeable in Fig.~\ref{fig:FigSRE4H}c when only considering the sum rules. However, the fact that the left feature in Fig.~\ref{fig:FigSRE4HTemp} shows a dip for all temperatures means that it should be related to a V scheme. Thus, the current assignment of the labels with corresponding pumping schemes is justified. Using equations~\ref{eq:L1} through \ref{eq:L4} (Suppl. Inf.), the effective g-factors can be determined. Fitting these to equation (\ref{eq:geff}) gives the values for $g_\parallel$ and $g_\perp$ reported in the main text.

\clearpage

\section{Franck-Condon principle with respect to spin}\label{sec:FC}

The amplitude of the two-laser emission signatures is determined by the strength of the underlying optical transitions. For a transition $\ket{g_i}$-$\ket{e_j}$, this strength is determined by the spin overlap $\bra{g_i}\ket{e_j}$, according to the Franck-Condon principle with respect to spin\cite{fox2001SI}. The quantum states of the spin in the electronic ground and excited state can be described using effective spin Hamiltonian
\begin{equation}\label{eq:spinHam2}
	H_{g(e)} = \mu_B \mathbf{B} \cdot \mathbf{g}_{g(e)} \cdot \tilde{\mathbf{S}}
\end{equation}
with $\mu_B$ the Bohr magneton, $\mathbf{B}$ the applied magnetic field vector,  $\tilde{\mathbf{S}}$ the effective spin vector, and where the ground (excited) state g-parameter is a tensor $\mathbf{g}_{g(e)}$. Using Cartesian coordinates this can be written as
\begin{equation}\label{eq:gtensor}
	\mathbf{g}_{g(e)} = \begin{pmatrix}
		g^{g(e)}_\perp & 0  & 0 \\
		0 & g^{g(e)}_\perp & 0 \\
		0 & 0 & g^{g(e)}_\parallel
	\end{pmatrix}
\end{equation}
Here the $z$-axis is parallel to the SiC c-axis, and the $x$ and $y$-axes lay in the plane perpendicular to the c-axis. Due to the symmetry of the defect, the magnetic field $\mathbf{B}$ can be written as
\begin{equation}
	\mathbf{B} = \begin{pmatrix}
		0 \\
		B\sin{\phi}\\
		B\cos{\phi}
	\end{pmatrix}
\end{equation}
where $B$ indicates the magnitude of the magnetic field. The resulting Hamiltonian $H_{g(e)}$ may be found by substituting $\textbf{B}$ and $\mathbf{g}_{g(e)}$ into equation (\ref{eq:spinHam2}), and considering that $S = 1/2$.
The basis of $H_{g(e)}$ can be found from the eigenvectors.

For the ground state $g^g_\perp$ is zero, thus the bases of $H_{g}$ and $S_z$ coincide, independent of $\phi$. Therefore, there is no mixing of spins in the ground state. However, in the excited state $g^e_\perp$ is nonzero, causing its eigenbasis to rotate if a magnetic field is applied non-parallel to the c-axis. The new eigenbasis is a linear combination of eigenstates of $S_x$, $S_y$ and $S_z$, such that there will be mixing for spins in the excited state for any nonzero angle $\phi$.

\begin{figure}[h!]
	\centering
	\includegraphics[width=10cm]{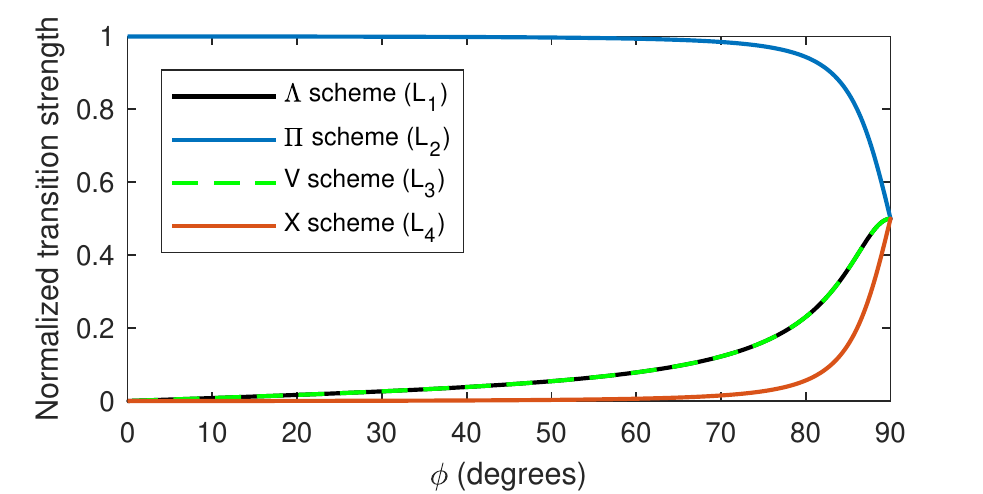}
	\caption{\textbf{Two-laser pumping scheme transition strengths.} For each scheme the product of the spin overlaps from both underlying transitions is shown. The strength of the $\Pi$ scheme is near unity for large angles and never vanishes. The strengths of the $\Lambda$ and V schemes are equal, they vanish at $\phi=\ang{0}$. The X scheme strength vanishes more rapidly than any other scheme for angles $\phi$ close to $\ang{0}$.}\label{fig:FigStrengths}
\end{figure}

We calculate the spin overlap for the $\ket{g_i}$-$\ket{e_j}$ transition from the inner product of two basis states $\ket{g_i}$ and $\ket{e_j}$. The strength of a two-laser pumping scheme is then the product of the strength of both transitions. For example, the strength of the $\Lambda$ scheme from Fig.~\ref{fig:FigSchemes}a equals the inner product $\bra{g_1}\ket{e_2}$ multiplied by $\bra{g_2}\ket{e_2}$. The resulting strengths for all four pumping schemes are depicted in Fig.~\ref{fig:FigStrengths}.

%%%%%%%%%%%%%%%%%%%%%%%%%%%%%%%%%%%%%%%
%%%%%%%%%%%%%%% Figure S7 %%%%%%%%%%%%%
%%%%%%%%%%%%%%%%%%%%%%%%%%%%%%%%%%%%%%%

We now compare these transition strengths to the data in Fig.~\ref{fig:FigSRE}b,c and Fig.~\ref{fig:FigMoreSRE} and \ref{fig:FigSRE4H}. It is clear that the $\Pi$ scheme is the strongest pumping scheme for all angles $\phi\neq \ang{90}$. This explains the large relative amplitude of $L_2$ in our measurements. The $\Lambda$ and V scheme transition strengths are equal, starting from zero for $\phi = \ang{0}$ and increasing as $\phi$ approaches $\ang{90}$. For the $\Lambda$ scheme, this is consistent with the increasing relative amplitude of $L_1$. For $\phi$ close to $\ang{90}$ the amplitude of $L_1$ is even larger than for $L_2$. The reason for this is that a $\Lambda$ scheme is emitting more effectively than a $\Pi$ scheme. The V scheme is harder to observe in the background emission, such that $L_3$ is only visible for $\phi$ close to $\ang{90}$. Finally, the transition strength of the X scheme is only significant for $\phi$ close to $\ang{90}$, which is why we have not been able to observe $L_4$ below $\ang{81}$ in 6H-SiC.

\clearpage

\section{V-scheme dip}\label{sec:SI-dip}
Understanding the observation of a dip for the V pumping scheme in a four-level system (Fig.~\ref{fig:FigSRE}c in the main text) is less trivial than for the observation of peaks from the other three pumping schemes. The latter can be readily understood from the fact that for proper two-laser detuning values both ground states are addressed simultaneously, such that there is no optical pumping into dark states. In this section we will investigate how a dip feature can occur in the PLE signals. Our modeling will be based on solving a master equation in Lindblad form with a density matrix in rotating wave approximation for a four-level system with two near-resonant lasers\cite{fleischhauer2005SI}.

\begin{figure}[h!]
	\centering
	\includegraphics[width=17cm]{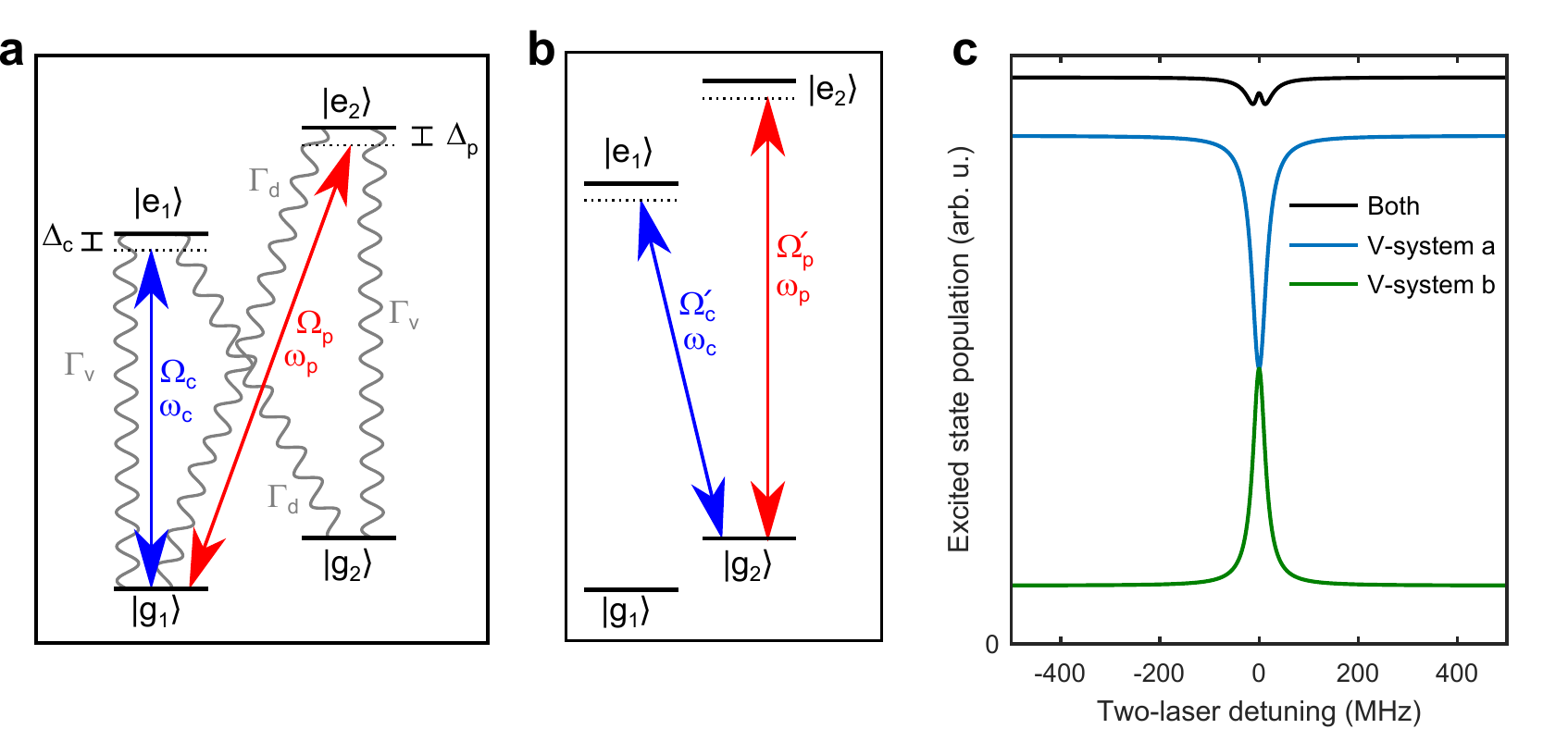}
	\caption{\textbf{Four-level V-scheme model.} \textbf{a,} V pumping scheme in a four level system. Here $\Omega$ is the Rabi frequency for the control and probe lasers, and $\omega$ their (angular) frequency. $\Gamma_v$ and $\Gamma_d$ are the decay rates for vertical and diagonal decay, respectively. $\Delta$ represents the detuning from resonance of the control and probe beam. \textbf{b,} V-scheme simultaneously resonant (with the scheme in panel \textbf{a}) for another part of the inhomogeneously broadened ensemble. Probe and control Rabi frequencies $\Omega'$ differ from \textbf{a}, since both lasers drive other transitions with different dipole strengths. \textbf{c,} Total population in the excited-state levels ($\ket{e_1}$ and $\ket{e_2}$) for both schemes separately (blue and green) as well as their sum (black).}
	\label{fig:Fig4level}
\end{figure}

\clearpage

Consider the four-level system depicted in Fig.~\ref{fig:Fig4level}a. A control laser is near-resonant with the $\ket{g_1}$-$\ket{e_1}$ (vertical) transition and a probe laser near-resonant with $\ket{g_1}$-$\ket{e_2}$ (diagonal) transition. Here the two-laser detuning is defined as $\delta = \Delta_p - \Delta_c$, \textit{i.e.}~the difference between the detunings $\Delta$ of both lasers from their respective near-resonant transitions, such that the emission feature appears at zero two-laser detuning. The decay rates from the excited states are $\Gamma_v$ and $\Gamma_d$ for vertical and diagonal transitions, respectively. They are quadratically proportional to the spin-state overlap $\braket{g_i}{e_j}$
\begin{align}
	\Gamma_v &\propto \left|\braket{g_1}{e_1}\right|^2,\label{eq:Gammav}\\
	\Gamma_d &\propto \left|\braket{g_1}{e_2}\right|^2.\label{eq:Gammad}
\end{align}
These rates are unequal, since the spin-state overlap for diagonal transitions is generally smaller than for vertical transitions (see previous section). The decay rates $\Gamma_e$ between excited-state levels  and $\Gamma_g$ ground-state levels  are assumed very small compared to the decay rates from the excited-state levels. The decay rates from ground-state levels towards the excited-state levels are set to zero. Dephasing rates are taken relative to the $\ket{g_1}$ state ($\gamma_{g1} = 0$). The choices for parameters are listed in table~\ref{tab:VschemeParam}. The Rabi frequencies $\Omega_c$ and $\Omega_p$ of the driven transitions are linearly proportional to the spin-state overlap
\begin{align}
	\Omega_c &\propto \left|\braket{g_1}{e_1}\right|,\label{eq:Omegac}\\
	\Omega_p &\propto \left|\braket{g_1}{e_2}\right|.\label{eq:Omegap}
\end{align}

\begin{table}
	\caption{\textbf{Parameter choices for V-scheme model}}
	\label{tab:VschemeParam}
	\begin{center}
		\begin{tabular}{ c@{\hskip 10mm}|c||c@{\hskip 10mm}|c } %@{\hskip 10mm}
			parameter & value (Hz) & parameter & value (Hz)\\\hline
			$\Gamma_v$ & $0.9\cdot10^7$ & $\gamma_{g1}$ & $0$\\
			$\Gamma_d$ & $0.1\cdot10^7$ & $\gamma_{g2}$ & $5\cdot10^6$\\
			$\Gamma_g$ & $1\cdot10^4$ & $\gamma_{g3}$ & $5\cdot10^6$\\
			$\Gamma_e$ & $1\cdot10^4$ & $\gamma_{g4}$ & $5\cdot10^6$\\
			$\Delta_c$ & $0$ & $\Omega_c$ & $\sqrt{.9}\cdot10^7$\\
			$\Delta_p$ & $\in [-500, 500]\cdot10^6$ & $\Omega_p$ & $\sqrt{.1}\cdot10^7$\\
		\end{tabular}
	\end{center}
\end{table}

Additionally, we have to consider a secondary V-scheme (Fig.~\ref{fig:Fig4level}b) resonant with another part of the inhomogeneously broadened ensemble. The control and probe laser are swapped, as the former now addresses a diagonal transition, while the latter addresses a vertical one. The new Rabi frequency is taken to be $\Omega_c' = \sqrt{\frac{\Gamma_d}{\Gamma_v}}\Omega_c$ for the control beam, which is now driving a diagonal transition (with reduced strength). The probe beam is driving a vertical transition (with increased strength), and its Rabi frequency is $\Omega_p' =\sqrt{\frac{\Gamma_v}{\Gamma_d}}\Omega_p$.

Considering both V-schemes, we calculate the total population in both excited-state levels as it reflects the amount of photoluminescence resulting from decay back to the ground states. The two-laser detuning dependence of the excited-state population is shown in Fig.~\ref{fig:Fig4level}c. The black curve considers both schemes simultaneously, which represents the situation in our measurements.
Here the dip indeed appears, although both separate schemes (a and b) display a dip and peak (respectively). The competition between both schemes limits the depth of the observed dip, which explains our observation of shallow dips in contrast to sharp peaks in Fig.~\ref{fig:FigSRE}c in the main text.

Interestingly, the black curve displays a peak within the dip, which might seem like a CPT feature. However, this feature is not visible in either curve from the two separate pumping schemes. This peak appears because the peak from the second V-scheme (green) is slightly sharper than the dip from the first one (blue). The peak might still be caused by CPT, as the blunting of the dip relative to the peak can be caused by a long dephasing time of the ground state. %This would make the experimental observation of CPT (in emission) for a V-scheme far less trivial compared to a $\Lambda$-scheme in a four-level system.

Key to understanding the appearance of a dip in the total photoluminescence emission is the difference in decay rates, vertical decay being favored over diagonal decay. Consider the pumping scheme from Fig.~\ref{fig:Fig4level}a. When the probe laser is off-resonant the control laser drives the $\ket{g_1}$-$\ket{e_1}$ transition. Decay will occur mostly towards the $\ket{g_1}$ state and occasionally to the dark $\ket{g_2}$ state. If the probe laser becomes resonant with the $\ket{g_1}$-$\ket{e_2}$ transition, the increased population in the $\ket{e_2}$ state will prefer to decay towards the dark $\ket{g_2}$ state. The overall decay towards the dark state is now increased. The secondary pumping scheme (Fig.~\ref{fig:Fig4level}b) works the other way around, where the diagonal transition is always driven by the control beam and a resonant probe beam will counteract some of the pumping into the dark state (now $\ket{g_1}$). However, the slightly increased emission from scheme b cannot fully counteract the decreased emission from scheme a (even when $\Omega_p = \Omega_c = \Omega_p'=\Omega_c'$).

\clearpage

\section{Modeling of coherent population trapping}\label{sec:CPTmodel}

For fitting the CPT traces in Fig.~\ref{fig:FigCPT} in the main text, we use a standard CPT description\cite{fleischhauer2005SI}, extended for strong inhomogeneous broadening of the optical transitions, and an approach similar to the one from the previous section. However (as compared to the previous section), the behavior of CPT has a more pronounced dependence on parameters, such that almost no assumptions have to be made. When taking the spin Hamiltonians as established input (section 4), the only assumption made is that the spin relaxation time in the ground state and excited state is much slower than all other decay process. This allows for setting up fitting of the CPT traces with only two free fit parameters, which correspond to the optical lifetime and the inhomogeneous dephasing time $T_2^*$.

Since two lasers couple both ground-state levels to a single common excited-state level, the other excited-state level will be empty. Therefore, we may describe this situation with a three-level system, where the PL is directly proportional to the excited-state population. The decay rates and Rabi frequencies are proportional to the Franck-Condon factors for spin-state overlaps $\braket{g_i}{e}$ in the same way as before (equations (\ref{eq:Gammav})-(\ref{eq:Omegap})). At this angle ($\phi = \ang{102}$) we calculate these factors to be
\begin{align}
	\braket{g_1}{e} & =0.9793\\
	\braket{g_2}{e} & =0.2022
\end{align}
according to the reasoning in section~\ref{sec:FC}. We take that the $\ket{g_1}$-$\ket{e}$ is a vertical transition and $\ket{g_2}$-$\ket{e}$ a diagonal one.

In order to account for inhomogeneous broadening throughout the ensemble,
the solution of the master equation is computed for a set of control-laser detunings $\Delta_c$ (see Fig.~\ref{fig:Fig4level}a) around zero, its range extending far beyond the two-laser detuning values $\delta$ (since we experimentally observed an inhomogeneous broadening much in excess of the spin splittings). In this case the probe-laser detuning becomes $\Delta_p = \Delta_c + \delta$. The resulting excited-state populations are integrated along the inhomogeneous broadening $\Delta_c$ (up to the point where the signal contribution vanishes) to give the PL emission as a function of two-laser detuning $\delta$. Analogous to the previous section, we have to consider a secondary $\Lambda$-scheme in order to fully account for the inhomogeneous broadening. The total PL emission is found by adding together the excited-state populations from both schemes.

We fit this model to the data presented in Fig.~\ref{fig:FigCPT} after subtracting a static background. We extract the inhomogeneous dephasing time $T_2^* = 0.32 \pm 0.08$~$\mu$s and an optical lifetime of $56 \pm 8$~ns. The errors are estimated from the spread in extracted dephasing times and lifetimes throughout the data sets.

\clearpage

\section{Anisotropic ${\rm {\bf g}}$-factor in the effective spin-Hamiltonian}\label{sec:gAnis}
\subsection{Relationship between effective spin Hamiltonian and local configuration of the defect}

An effective spin-Hamiltonian as the one used in the main text is a convenient tool which allows us to describe the behavior of the system in a wide range of configurations, as long as the effective parameters are experimentally determined and all relevant states are considered. It is often the meeting point between experimentalists, who measure the relevant parameters, and theoreticians, who attempt to correlate them to the Hamiltonian that describes the configuration of the system. A careful description of the latter, and how it modifies the parameters at hand, allows us to rationalize our choices when investigating defects with varying characteristics (such as a different charge state or element). This task is more approachable when we consider the group-theoretical properties of the system at hand. Here we combine group-theoretical considerations with ligand field theory in order to qualitatively describe the features observed in our experiment. In particular, we aim at explaining the large Zeeman splitting anisotropy observed in both ground and excited states, and correlating it to the charge and spatial configuration of the defect.

In our experiments, we observe a single zero-phonon line (ZPL) associated with optical transitions between two Kramers doublets (KD, doublets whose degeneracy is protected by time-reversal symmetry and is thus broken in the presence of a magnetic field) in defects which contain Mo. The presence of a single zero-phonon line in both 4H and 6H-SiC samples indicates that the defect occupies a lattice site with hexagonal symmetry. The lattice of 6H-SiC has two inequivalent sites with cubic symmetry. Thus, if the defect were to occupy sites of cubic symmetry, we would expect to observe two ZPLs closely spaced in this sample. The absence of the ZPL associated with this defect in samples of 3C-SiC\cite{ivady2012SI} further corroborates this assumption. Additionally, we observe strong anisotropy in the Zeeman splitting of the ground and excited states. Specifically, when the magnetic field is perpendicular to the symmetry axis of the crystal, the Zeeman splitting of the ground state goes to zero, whereas that of the excited state is very small. This feature is observed in other transition-metal defects in SiC situated at Si substitutional sites of hexagonal symmetry and with one electron in its $3$d orbital\cite{baur1997SI}, but we are not aware of a clear explanation of the phenomenon.

In our experiments, we observed transitions between sublevels of doubly degenerate ground and excited states, whose degeneracy is broken in the presence of a magnetic field. Thus, we note that ground and excited states are isolated KDs, indicating that the defect contains an odd number of electrons. A Mo atom has $6$ electrons in its valence shell. The atom can occupy a Si substitutional site ($\rm Mo_{Si}$), where it needs to bond to $4$ neighboring atoms, or an asymmetric split vacancy (ASV) site ($\rm Mo_{V_{Si}-V_{C}}$), where it bonds to $6$ neighboring atoms. These defects can, respectively, be described by a Mo ion in the configurations $\rm 4d^2$ and $\rm 4d^0$, indicating that the defect must be ionized in order to contain an odd number of electrons. Its charge state, which could be $\pm1$, $\pm3$, \textit{etc.}, is determined by the Fermi level in the crystal of interest. We note that the ZPL could only be observed in p-doped samples, which indicates that the features investigated here are unlikely to arise from negatively charged defect. The defect $\rm Mo_{Si}^{+1}$ (where $+1$ represents the charge state of the defect, not the Mo atom) can be approximately described by a Mo in a configuration $\rm 4d^1$, which facilitates the treatment of its configuration in terms of d orbitals. In contrast, the defect $\rm Mo_{V_{Si}-V_{C}}^{+1}$ is described by an electronic configuration containing a hole in the bonding orbitals. These orbitals show strong hybridization between the d orbitals of the Mo and the orbitals of the ligands, and cannot be straight-forwardly analyzed using the formalism described below. Nonetheless, inspired by the similarities between our system and other transition-metal defects reported in SiC\cite{baur1997SI}, we investigate the effect of the crystal field of $\rm C_{3v}$ symmetry --which is expected to be significant in hexagonal lattice sites in 4H-SiC and 6H-SiC-- on the one-electron levels of the $5$ sublevels ($10$, if spin multiplicity is included) of the $4$d shell of a Mo atom. We qualitatively predict the spin-hamiltonian parameters expected for a Mo ion in a $\rm 4d^1$ configuration, and compare our analysis to the experimental results.

\subsection{Ion in $\rm 4d^1$ configuration in the presence of crystal field of $\rm C_{3v}$ symmetry and spin-orbit coupling}

\begin{figure}[h!]
	\centering
	\includegraphics[width=7cm]{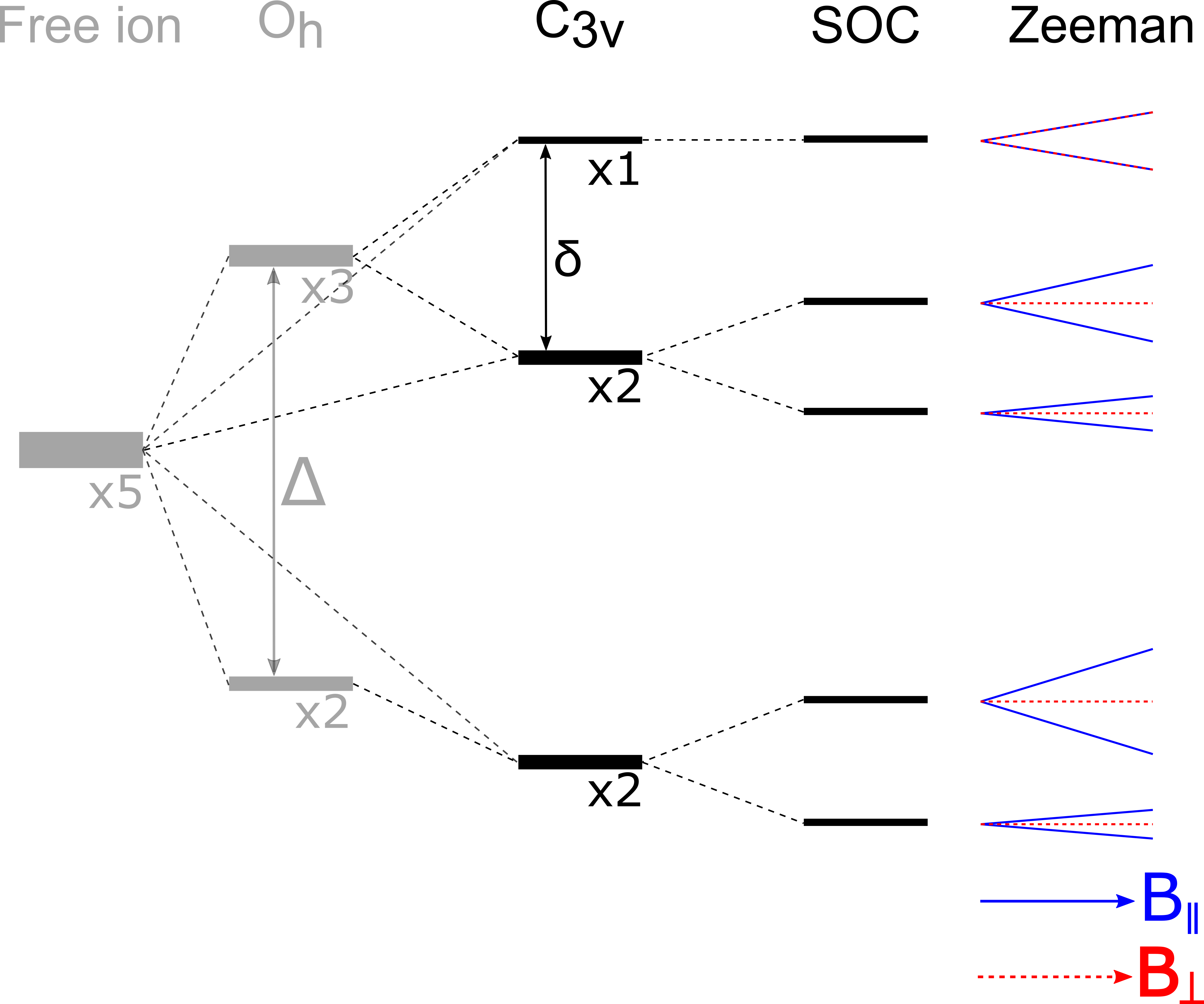}
	\caption{\textbf{Splitting of one-electron energy levels of a $4$d orbital}, under the action of a crystal field and spin-orbit coupling. In the free atom, the $5$ orbitals corresponding to the $4$d shell (disregarding the spin) are degenerate. A crystal field of cubic symmetry breaks this degeneracy,  generating an orbital triplet and a doublet, whereas a crystal field of $\rm C_{3v}$ symmetry, splits the $5$ orbitals into one singlet and two doublets. In the text, we focus on a crystal field of $\rm C_{3v}$ symmetry, and disregard the cubic term. Although we recognize that this is an approximation, we argue that this approach clarifies the physics governing the strong magnetic anisotropy observed, and is thus justified. Spin-orbit coupling is responsible for splitting the doublets, generating in total $5$ sets of Kramers doublets (here, the spin of the electron is taken into account). The energy splittings caused by a magnetic field within these KD give rise to the effective spin Hamiltonian parameters considered. We note that a group-theoretical approach alone is not capable of providing the order of the energy levels shown in the figure. We take this order to be the one observed in transition-metal defects in a tetrahedral crystal field with strong trigonal distortion\cite{abragam1970SI}.}
	\label{fig:Perturbation}
\end{figure}

The $5$ degenerate sublevels of a $4$d-orbital are split by a crystal field of $\rm C_{3v}$ symmetry\cite{abragam1970SI}. The energy splittings induced by this field are much smaller than the energy difference between the $4$d shell and the next orbital excited state ($5$s). This allows us to, initially, consider the $5$ orbitals of the $4$d shell as a complete set. Since Mo is a heavy atom, we cannot disregard the effect of spin-orbit interaction. However, we assume that the crystal field is larger than the effect of SOC, that is, $\Delta E_{free} \gg \Delta E_{crystal} \gg \Delta E_{spin-orbit} \gg \Delta E_{Zeeman}$, where $\Delta E$ denotes the energy splitting induced by each term (see Fig.~\ref{fig:Perturbation}).

The $5$ orbital states of the d-orbital form a 5-dimensional irreducible representation (irrep) of the full rotation group SO($3$). When the symmetry is lowered by the crystal field to $\rm C_{3v}$, the $5$-dimensional representation is split into $2$ doublets ($E_1$, $E_2$) and $1$ singlet ($A$) that are irreps of $\rm C_{3v}$. Writing the $5$ components of the $4$d orbital in terms of the quadratic functions $z^2$, $x^2 - y^2$, $xy$, $xz$, $yz$ allows us to identify which orbitals are degenerate in the presence of a crystal field of trigonal symmetry. We find that the singlet $A$ is composed of the orbital $\rm 4d_{z^2}$. Furthermore, the orbitals $\rm 4d_{xz}$ and $\rm 4d_{yz}$ are degenerate upon action of the crystal field and make up doublet $E_1$. Finally the orbitals $\rm 4d_{x^2 - y^2}$ and $\rm 4d_{xy}$ correspond to doublet $E_2$. Group-theoretical considerations alone are not capable of elucidating which irrep corresponds to the ground state, that is, it does not provide information about the order of the energy levels.

Comparison between the Cartesian form of these $5$ orbitals and the spherical harmonics which span a $5$-dimensional space (that is, the spherical harmonics $Y^m_l$ with $l = 2$) allows us to rewrite the relevant orbitals as linear combinations of the eigenstates of the operators $L^2$, $L_z$. This yields a new basis for each irrep considered above:
\begin{align}
	&E_1: Y^{-2}_2 = \ket{d_{-2}}; Y^2_2 = \ket{d_{2}} &\text{1st orbital doublet}\\
	&E_2: Y^{-1}_2 = \ket{d_{-1}}; Y^{1}_2 = \ket{d_1} &\text{2nd orbital doublet}\\
	&A: Y^0_2 = \ket{d_0} &\text{orbital singlet}
\end{align}
\noindent When the spin multiplicity is considered, each orbital doublet yields $4$ possible states, whereas the orbital singlet yields $2$ possible states. Spin-orbit coupling (represented by the operator $H_{SO} = -\lambda \mathbf{L} \cdot \mathbf{S}$) is responsible for splitting these states into 5 different Kramers doublets:
\begin{align}
	&\text{KD1}: \ket{d_{+2}, +\tfrac{1}{2}}; \ket{d_{-2}, -\tfrac{1}{2}}\label{eq:Basis1}\\
	&\text{KD2}: \ket{d_{+2}, -\tfrac{1}{2}}; \ket{d_{-2}, +\tfrac{1}{2}}\label{eq:Basis2}\\
	&\text{KD3}: \ket{d_{+1}, +\tfrac{1}{2}}; \ket{d_{-1}, -\tfrac{1}{2}}\label{eq:Basis3}\\
	&\text{KD4}: \ket{d_{+1}, -\tfrac{1}{2}}; \ket{d_{-1}, +\tfrac{1}{2}}\label{eq:Basis4}\\
	&\text{KD5}: \ket{d_{0}, +\tfrac{1}{2}}; \ket{d_{0}, -\tfrac{1}{2}}\label{eq:Basis5}
\end{align}
\noindent where the basis vectors are given in terms of the quantum numbers $m_l$ and $m_s$ which denote the projection of the orbital and spin angular momentum along the quantization axis, respectively (Fig.~\ref{fig:Perturbation}). Here, the spin-orbit coupling is considered up to first order in the energy correction, whereas the wave function is corrected up to zeroth order.

A magnetic field lifts the degeneracy between the two components of each KD. This splitting is usually described phenomenologically by an effective Zeeman Hamiltonian in a system with pseudospin $\mathbf{\tilde{S}} = \tfrac{1}{2}$.
\begin{align}
	H_{eff} =& -\mu_B \mathbf{B} \cdot \mathbf{g} \cdot \mathbf{\tilde{S}}_{1/2}
	\label{eq:Zeeman_p}
\end{align}
\noindent where $\mu_B$ is the Bohr magneton, $\mathbf{B}$ the magnetic field vector, $\mathbf{\tilde{S}}_{1/2}$ the pseudo spin $\tfrac{1}{2}$ operator and $\mathbf{g}$ the g-tensor. In the presence of axial symmetry, $\mathbf{g}$ can be diagonalized such that equation (\ref{eq:Zeeman_p}) can be rewritten in terms of the symmetry axis of the crystal
\begin{align}
	H_{eff} =& -\mu_B \Big(g_\parallel B_z  \tilde{S}_{1/2,z} + (g_\bot B_x  \tilde{S}_{1/2,x} + g_\bot B_y  \tilde{S}_{1/2,y})\Big)
	\label{eq:Zeeman_pII}
\end{align}

In terms of the eigenstates belonging to each KD, the splitting is described by the Zeeman Hamiltonian given by
\begin{align}
	H_{Zee} = -\mathbf{B} \cdot \boldsymbol{\mu} = -\mu_B \mathbf{B} \cdot (g_0 \mathbf{S} + k\mathbf{L}) \label{eq:Zeeman}
\end{align}
\noindent where $\boldsymbol{\mu}$ is the magnetic moment operator, $g_0$ the g-factor for a free electron, $\mathbf{S}$ the total spin operator, $k$ the orbital reduction factor, and $\mathbf{L}$ the orbital angular momentum operator\cite{abragam1970SI,chibotaru2005SI}. The orbital reduction factor $k$, is a factor between $0$ and $1$ which corrects for partial covalent bonding between the electron and the ligands\cite{abragam1970SI} (note that the value of $k$ differs for each of the 5 KDs in equations~(\ref{eq:Basis1}-\ref{eq:Basis5})). Comparison of equations (\ref{eq:Zeeman_pII}) and (\ref{eq:Zeeman}) shows that
\begin{align}
	g_\parallel =& 2 \expval{g_e S_z + k L_z} = \frac{2 \expval{\mu_z}}{\mu_B}\label{eq:gpar}\\
	g_\bot =& 2 \expval{g_e (S_x+S_y) + k (L_x+L_y)} = \frac{2 \expval{\mu_x + \mu_y}}{\mu_B} \label{eq:gbot}
\end{align}

As long as the magnitude of this Zeeman splitting is small compared to the spin-orbit interaction we can consider, to first order, the effect of the magnetic field in the sets formed by each KD independently. That is, we consider that the magnetic field does not mix states pertaining to two different KDs.

In order to calculate the values of $g_\parallel$ and $g_\bot$ for each KD defined by trigonal symmetry and spin-orbit coupling, we rewrite equation (\ref{eq:Zeeman}) as
\begin{align}
	H_{Zee} = - (B_z \mu_z + B_x \mu_x + B_y \mu_y ) = - \big(B_z \mu_z + \tfrac{1}{2}( B_+ \mu_- + B_- \mu_+)\big)
\end{align}
\noindent where the $+$ and $-$ subindices denote the raising and lowering magnetic moment operators and the linear combinations $B_x \pm i B_y$, respectively. When we consider the basis given in equations (\ref{eq:Basis1}-\ref{eq:Basis4}), the matrix elements of both $\mu_+$ and $\mu_-$ are zero between two eigenvectors pertaining to one KD. This arises from the fact that the operator $\mu_+$ couples states with $(m_l, m_s)$ to states with $(m_l + 1, m_s)$ or $(m_l, m_s+1)$. Since, within a KD, there is a change in both $m_l$ and $m_s$ when going from one eigenvector to the other, the operators $\mu_+$ and $\mu_-$ cannot couple these states to each other. Explicitly, for KD1 for example, we obtain
\begin{align}
	\mel{d_{+2}, -\tfrac{1}{2}}{\mu_\pm}{d_{+2}, -\tfrac{1}{2}} = 0\\
	\mel{d_{+2}, -\tfrac{1}{2}}{\mu_\pm}{d_{-2}, +\tfrac{1}{2}} = 0\\
	\mel{d_{-2}, +\tfrac{1}{2}}{\mu_\pm}{d_{-2}, +\tfrac{1}{2}} = 0
\end{align}
\noindent and in a similar way for KDs 2 through 4. Thus, up to first order, a magnetic field applied perpendicular to the crystal c-axis is not capable of lifting the degeneracies of the 4~KDs given in equations (\ref{eq:Basis1})-(\ref{eq:Basis4}). Comparing these results to equation (\ref{eq:gbot}) we conclude that, for the $8$ sublevels of the KDs 1~through 4, $g_{\bot} = 0$. This arises from the effect of both the crystal field of C$_{\rm 3v}$ symmetry and SOC in decoupling and isolating KDs with the properties mentioned above. This is not the case for KD5, given in equation (\ref{eq:Basis5}). In this case,
\begin{align}
	\mel{d_{0}, -\tfrac{1}{2}}{\mu_\pm}{d_{0}, +\tfrac{1}{2}} \neq 0
\end{align}
\noindent and the degeneracy of this KD is broken in the presence of a magnetic field perpendicular to the c-axis of the crystal.

We can consider in addition the effect of spin-orbit coupling in mixing the eigenstates presented in equations (\ref{eq:Basis1}-\ref{eq:Basis5}). Spin-orbit coupling is responsible for mixing between the eigenstates of KD2 and KD3 (equations (\ref{eq:Basis2}) and (\ref{eq:Basis3})). Since both of these KDs show $g_\bot = 0$, this mixing does not modify the expected value of $g_\bot$ in neither KD. In contrast, the SOC induced mixing between KD4 and KD5 causes some deviation of $g_\bot$ from $0$ in KD4, since $g_\bot \neq 0$ in KD5. The values of $g_\parallel$ and $g_\bot$ for one electron in each of the KDs described in this section are presented in table~\ref{tab:Table_g}.

From the 5 KDs in equations~(\ref{eq:Basis1}-\ref{eq:Basis5}), one KD is the ground state and one KD is the excited state that we address in our experiments.
As said before, our group-theoretical approach cannot identify the ordering in energy of these 5 KDs. However, by looking at the g-factor properties of the KDs in table~\ref{tab:Table_g} we can check which ones show consistent behavior with that of the observed ground and excited state. For the observed ground state, we found $|g_\parallel| < 2$ and $g_\bot = 0$. Concerning possible values for the orbital reduction factor $k$, by definition $k < 1$, and we must have $k > 0.1$ since $|g_\parallel|$ deviates substantially from 2. This suggests that KD2 is the ground state. For the excited state, we also have $|g_\parallel| < 2$, but with $g_\bot \gtrapprox 0$. This suggests KD4 is the excited state we observed in our experiments.
In addition, the optical transition observed is mainly polarized along the crystal c-axis of the defect. Careful analysis of the selection rules associated with the double trigonal group (which includes, besides the spatial symmetry, the spin of the electron) has been reported by Kunzer \textit{et al.}\cite{kunzer1993SI}. Comparing their results to the considerations presented in the previous paragraphs confirms that the transition between KD2 and KD4 is predominantly polarized along the crystal c-axis, as observed. Finally, we note that we could not experimentally identify secondary ZPLs corresponding to transitions between other sets of KDs, even though they are allowed by symmetry. This could be explained by a series of factors. On the one hand, some of the KDs treated could have energies above the conduction band edge in the crystal, which would impede the observation of optical transitions from and into these levels. On the other hand, the presence of these lines could be masked by the intense phonon sideband at the red side of the ZPL, or the associated photon energies fall outside our detection window.

\begin{table}
	\caption{The g-factors of the Kramers doublets originated due to spin-orbit coupling within each subspace of the electronic eigenstates in a field of  $\rm C_{3v}$ symmetry. Spin-orbit coupling is added as a perturbation, and included up to first order. The parameters $\lambda$ and $\delta$ are as defined in the text and in Fig.~\ref{fig:Perturbation}. Note that the g-factor values in this table can take on negative values, while in our experimental analysis we can only extract $|g_\parallel|$ and $|g_\bot|$.}
	\label{tab:Table_g}
	\begin{center}
		\begin{tabular}{ |l|l|l|l| }
			\hline
			$\rm C_{3v}$ & Spin-Orbit  & $g_\parallel$ & $g_\bot$\\ \hline
			\multirow{2}{*}{Doublet, $m_l = \pm 2$} & KD1, eq. \ref{eq:Basis1} & $2(2k+1)$  & 0  \\
			& KD2, eq. \ref{eq:Basis2} & $2(2k-1)$  & 0 \\ \hline
			\multirow{2}{*}{Doublet, $m_l = \pm 1$} &  KD3, eq. \ref{eq:Basis3} & $2(k+1)$ & 0  \\
			&  KD4, eq. \ref{eq:Basis4} & $2(k-1)$ & $0~+ \propto \tfrac{\lambda}{\delta}$ \\ \hline
			Singlet, $m_l = 0$ &  KD5, eq. \ref{eq:Basis5} & 2 & $2~- \propto \tfrac{\lambda}{\delta}$\\ \hline
		\end{tabular}
	\end{center}
\end{table}

\subsection{Validity of our assumptions}

The model considered here is capable of qualitatively informing us about the behavior of orbitals with d character in the presence of trigonal crystal field and spin orbit coupling. It is clear that the full description of the configuration of the defect is far more subtle than the simple model applied here. We intend to comment on this in the next paragraphs.

\paragraph{Symmetry of the crystal field.} In our derivation, we assume that the trigonal crystal field is the prevailing term in the Hamiltonian describing the defect. This assumption is not rigorously correct, since the symmetry of defects in SiC is more accurately described by a ligand field of cubic symmetry -- which determine most of its ground and excited state properties. This field is modified in the presence of axial symmetry, as is the case for defects in hexagonal lattice sites, which is generally included as a first-order perturbation term in the Hamiltonian. Nonetheless, it can be shown\cite{abragam1970SI,dietz1963SI} that the large anisotropy in the Zeeman response described above, with the cancelation of $g_\bot$, is also observed in the case of a cubic field with trigonal distortion and spin-orbit coupling of similar magnitudes. The analysis, in this case, is more laborious due to the fact that mixing of the orbitals is involved, and calculating the matrix elements of the operators $L_\pm$, $S_\pm$, $L_z$ and $S_z$ is less trivial. Furthermore, this analysis would not increase our level of understanding of the system at this point, since we were only capable of observing transitions between the sublevels of two KDs in this experiment. This approach would be more profitable if transitions between other sets of KDs were observed, allowing us to unravel several parameters associated with the system, such as the strength of the spin-orbit coupling and trigonal crystal field.

\paragraph{Charge state of the defect.} Similarly, it can be shown that the considerations presented here can be expanded to configurations where the $4$d orbitals are filled by multiple electrons (for instance, a defect in a configuration $\rm 4d^3$). In this case, a doubly degenerate orbital configuration (in symmetry terms, a configuration of the kind $^m$E, where $m$ is the spin multiplicity) in the presence of a crystal field of $\rm C_{3v}$ symmetry gives rise to at least one KD with $g_\bot = 0$ when SOC is take into account. Nonetheless, only a negatively charged Mo in a Si substitutional site would give rise to a defect in the configuration $\rm 4d^3$. The absence of the ZPL in n-doped samples indicates that this is unlikely.

In addition, a similar group theoretical analysis can show that one hole in a bonding orbital of symmetry E would also give rise to $g_\bot = 0$. Thus, the features observed here could also correspond to a positively charged $\rm Mo_{V_{Si}-V_{C}}$ defect (where one of the six Mo electrons that participate in bonding is lost to the crystal lattice). Due to the strong hybridization between the Mo and the divacancy orbitals in this case, the description of this case is more subtle and will not be performed here.

\subsection{Summary}

We showed that an analysis of the effect of the defect symmetry on the Zeeman energy splittings of its ground and excited states, combined with the experimental observations, helps us unravel the configuration of the defect studied in this work. We show that, in $\rm C_{3v}$ symmetry, a combination of the crystal field and spin-orbit interaction is responsible for the strong magnetic anisotropy observed experimentally. Furthermore, the fact that the defect studied in this work is only observed optically in samples which are p-doped indicates that the charge of the defect is more likely positive than negative. In this way, we conclude that the most probable configuration of our defect is a Mo ion on a Si substitutional site of h symmetry, with a charge $+1$, which can be approximately described by a Mo atom in a $\rm 4d^1$ configuration. The absence of other lines associated with the defect prevents us from providing a more accurate description of the system. Nonetheless, we have developed a qualitative description based on symmetry, which explains the Zeeman splittings observed. The considerations presented here allow us to predict and rationalize the presence of strong anisotropy in other transition-metal defects in SiC. We expect neutrally charged vanadium defects in hexagonal lattice sites to show a magnetic behavior similar to the one observed in the Mo defects investigated in this work.

%\end{document}

\end{document}